\documentclass[11pt,a4paper]{article}

\RequirePackage{ifpdf} 
\usepackage{amsmath} 
\usepackage{mathtools}

\usepackage{jheppub}
\usepackage{pstricks}
\usepackage[final]{pdfpages} 
\usepackage{ifpdf} 
\usepackage{slashed}
\usepackage{hyperref}

\usepackage{color} 
\usepackage{graphics}

\usepackage{etoolbox} 
\usepackage{fixmath}

\usepackage{caption} 
\usepackage{subcaption} 
\usepackage{amsfonts}

\usepackage{multirow}
\usepackage{epstopdf}

\usepackage{relsize}
\usepackage{float}

\usepackage{tikz}
\usetikzlibrary{positioning,arrows}
\usetikzlibrary{decorations.pathmorphing}
\usetikzlibrary{decorations.markings}
\usetikzlibrary{shapes.geometric}
\usepackage{endnotes}
\tikzset{
    vector/.style={decorate, decoration={snake}, draw},
    provector/.style={decorate, decoration={snake,amplitude=2.5pt}, draw},
    antivector/.style={decorate, decoration={snake,amplitude=-2.5pt}, draw},
    fermion/.style={draw=black,
      postaction={decorate},decoration={markings,mark=at position .55
        with {\arrow[draw=black]{>}}}}, 
    fermionbar/.style={draw=black, postaction={decorate},
                       decoration={markings,mark=at position .55 with {\arrow[draw=black]{<}}}},
    fermionnoarrow/.style={draw=black},
    gluon/.style={decorate, draw=black,decoration={coil,amplitude=4pt, segment length=6pt}},
    scalar/.style={dashed,draw=black,
      postaction={decorate},decoration={markings,mark=at position .55
        with {\arrow[draw=black]{>}}}}, 
    scalarbar/.style={dashed,draw=black,
      postaction={decorate},decoration={markings,mark=at position .55
        with {\arrow[draw=black]{<}}}}, 
    scalarnoarrow/.style={dashed,draw=black},
    electron/.style={draw=black,
      postaction={decorate},decoration={markings,mark=at position .55
        with {\arrow[draw=black]{>}}}}, 
    bigvector/.style={decorate, decoration={snake,amplitude=4pt}, draw},
}

\title{Two loop QCD amplitudes for di-pseudo scalar production in
gluon fusion}

\author[a]{Arunima Bhattacharya,}
\author[a,b]{Maguni Mahakhud \footnote{Since August 2019 at IISER Mohali},}
\author[a]{Prakash Mathews,}
\author[c]{and V. Ravindran}

\affiliation[a]{Saha Institute of Nuclear Physics, HBNI, 1/AF Saltlake,
Kolkata 700064, India}
\affiliation[b]{Indian Institute of Science Education and Research Mohali, Knowledge city,
Sector 81, SAS Nagar, Manauli PO 140306, India}
\affiliation[c]{The Institute of Mathematical Sciences, HBNI, Taramani, Chennai-600113, India}
\emailAdd{arunima.bhattacharya@saha.ac.in}
\emailAdd{maguni@iisermohali.ac.in}
\emailAdd{prakash.mathews@saha.ac.in} 
\emailAdd{ravindra@imsc.res.in}

\abstract{We compute the radiative corrections to 
the four-point amplitude  $g+g \rightarrow A+A$ in massless Quantum Chromodynamics (QCD) 
up to order $a_s^4$ in perturbation theory.  We used the effective field theory that
describes the coupling of pseudo-scalars to gluons and quarks directly, in the large top
quark mass limit.  Due to the CP odd nature of the pseudo-scalar Higgs boson, the
computation involves careful treatment of chiral quantities in dimensional regularisation.
The ultraviolet finite results are shown to be consistent with the universal infrared
structure of QCD amplitudes.  The infrared finite part of these amplitudes constitutes
the important component of any next to next to leading order corrections to observables
involving pair of pseudo-scalars at the Large Hadron Collider.    
}
\begin{document}

\preprint{\hspace{12 cm} IMSc/2019/09/08}

\allowdisplaybreaks[4]
\unitlength1cm
\maketitle
\flushbottom

\def\D{{\cal D}}
\def\g{\overline {\cal G}}
\def\gm{\gamma}
\def\ep{\epsilon}
\def\zo{\overline{z}_1}
\def\zt{\overline{z}_2}
\def\zob{\overline{z}_1}
\def\ztb{\overline{z}_2}
\def\C{{C}}
\def\C{{C}}
\def\Aob{\overline A_1^I}
\def\Atb{\overline A_2^I}
\def\Athb{\overline A_3^I}
\def\Afb{\overline A_4^I}
\def\Ao{A_1^I}
\def\At{A_2^I}
\def\Ath{A_3^I}
\def\fo{f_1^I}
\def\ft{f_2^I}
\def\fth{f_3^I}
\def\Af{A_4^I}
\def\Bo{B_1^I}
\def\Bt{B_2^I}
\def\Bth{B_3^I}
\def\Dob{\overline D_1^I}
\def\Dobd{\overline D_{d,1}^I}
\def\Dtb{\overline D_2^I}
\def\Dtbd{\overline D_{d,2}^I}
\def\Dthb{\overline D_3^I}
\def\Dthbd{\overline D_{d,3}^I}
\def\btob{\overline \beta_1}
\def\bttb{\overline \beta_2}
\def\btthb{\overline \beta_3}
\def\lfr{\log\left({\mu_F^2 \over \mu_R^2}\right)}
\def\lfrt{\log^2\left({\mu_F^2 \over \mu_R^2}\right)}
\def\lfrtt{\log^3\left({\mu_F^2 \over \mu_R^2}\right)}
\def\lqr{\log\left({q^2 \over \mu_R^2}\right)}
\def\lqrt{\log^2\left({q^2 \over \mu_R^2}\right)}
\def\lqrtt{\log^3\left({q^2 \over \mu_R^2}\right)}
\def\lmt{\log\left({\mu_R^2 \over m_t^2}\right)}
\def\lmts{\log^2\left({\mu_R^2 \over m_t^2}\right)}
\def\lw{\ln(1-\omega)}
\def\w{\omega}
\def\one{\ln(w)}
\def\two{\ln^2(w)}
\def\three{\ln^3(w)}
\def\four{\ln^4(w)}
\def\five{\ln^5(w)}
\def\six{\ln^6(w)}
\def\aLqf{\log\left({q^2 \over \mu_F^2}\right)} 
\def\aLqftwo{\log^2\left({q^2 \over \mu_F^2}\right)} 
\def\aLqfthree{\log^3\left({q^2 \over \mu_F^2}\right)} 
\def\M{{\cal M}}
\def\ep{\epsilon}
\def\unM{\hat{\cal M}}
\def\unas{ \left( \frac{\hat{a}_s}{\mu^{\epsilon}} S_{\epsilon} \right) }
\def\rnM{{\cal M}}
\def\rnas{ \left( a_s  \right) }
\def\b0{\beta_0}
\def\cD{{\cal D}}
\def\cC{{\cal C}}
\def\ca{\text{\tiny C}_\text{\tiny A}}
\def\cf{\text{\tiny C}_\text{\tiny F}}

\def\spt{(s+t)}
\def\spu{(s+u)}
\def\tpu{(t+u)}

\let\footnote=\endnote
\renewcommand*{\thefootnote}{\fnsymbol{footnote}}


\section{Introduction}


The discovery of Higgs boson of the standard model (SM) 
by ATLAS \cite{Aad:2012tfa} and CMS \cite{Chatrchyan:2012xdj} collaborations of the
Large Hardron Collider (LHC) has not only put the SM on strong footing but also opened up 
a plethora of
opportunities to investigate its properties and coupling to other SM particles.  
In certain beyond the SM (BSM) scenarios, one has enlarged Higgs sector, which allows 
more than one Higgs boson
\cite{Fayet:1974pd,Fayet:1977yc,Dimopoulos:1981zb,Sakai:1981gr,Inoue:1982pi,
Inoue:1983pp,Inoue:1982ej}. For example, 
the minimal supersymmetric SM (MSSM), there are five Higgs bosons, out of which
two of them are neutral scalars (h,H), one of them a pseudo scalar (A) and the remaining two are 
charged scalars ($H^\pm$).  
The pseudo scalar Higgs boson  which is CP odd 
could be as light as the discovered Higgs boson.  Hence, a dedicated effort has been going on 
to determine the CP property of the discovered Higgs boson to identify  
with that of SM, which allows only CP even.  This requires precise predictions for relevant observables
for both scalar and pseudo scalar ones.  
In the case of SM Higgs boson, the production cross section has been computed in perturbative 
Quantum Chromodynamics (QCD) to unprecedented accuracy.  This is possible, thanks to the fact that
the top quark degrees of freedom can be integrated out.  This results in an effective field theory (EFT)
where scalar Higgs boson couples directly to the gluons even at leading order (LO).   
In the context of light Higgs boson in MSSM, unlike the SM one, the mass is calculable. 
In \cite{Martin:2007pg,Harlander:2008ju,Kant:2010tf}, 
higher order radiative corrections to the mass are obtained to very good accuracy.
For the pseudo scalar Higgs boson, there have been efforts to achieve precision in 
the predictions for production cross sections at the LHC.
In \cite{Plehn:1996wb},first results on the production rate at NLO level in QCD for the pseudo scalar at 
the hadron collider appeared.  This was done by keeping non-zero top quark mass. 
In \cite{Dawson:1998py} EFT framework was set up by integrating out top quark fields, which
opened up the possibility of obtaining observables beyond NLO level as 
there is a reduction of number of loops compared to those  in
the full theory.  Unlike the case of CP even Higgs boson, inclusive cross section for the 
production of pseudo scalar Higgs is known \cite{Harlander:2002vv,Anastasiou:2002wq,Ravindran:2003um} only up to next-to-next-to leading order (NNLO) in pQCD.  
For N$^3$LO predictions, one requires three loop virtual amplitudes and real emission 
contributions.   The computation of
virtual corrections is technically challenging \cite{Ahmed:2015qpa} 
as pseudo scalar Higgs boson couples to SM fields through
two composite operators that mix under renormalisation.  In addition, these operators involve Levi-Civita
tensor and $\gamma_5$ which are hard to define in dimensional regularisation.  The three loop form factor
thus obtained was later combined with appropriate soft distribution function 
\cite{Ravindran:2005vv,Ravindran:2006cg,Ahmed:2014cla} and 
mass factorisation kernels to
obtain soft plus virtual contribution at N$^3$LO in QCD \cite{Ahmed:2015qda}.  
Later, the process dependent resummation constants
from the three loop form factors were used to perform threshold resummation in \cite{Ahmed:2016otz} 
and also make approximate
prediction at N$^3$LO level.  This was possible due to the similarity of the interaction vertices 
of scalar and pseudo scalar Higgs bosons with the gluons. 

Recently, there have been a surge of interest to study the production of pair of Higgs bosons
to determine Higgs self coupling, whose strength is a prediction of the SM, if the mass of the
Higgs boson is known.  Measurement of this coupling will provide an independent test on nature
of the Higgs boson.  The gluon gluon fusion subprocess producing pair of Higgs bosons through a 
heavy quark loop \cite{Glover:1987nx,Plehn:1996wb} is the dominant one at the LHC, however 
the cross section is only few tens of fb, making it very difficult to observe.  QCD corrections
not only increase the cross section but also stabilise the predictions
against renormalisation $\mu_R$, and factorisation $\mu_F$  scales.
NLO QCD corrections \cite{Dawson:1998py} and later on the top quark mass effects
are systematically taken into account in \cite{Grigo:2013rya,Frederix:2014hta,Maltoni:2014eza,Degrassi:2016vss,
Borowka:2016ehy, Borowka:2016ypz}.  Beyond NLO, an EFT where top quark degrees of freedom
are integrated out is used.  At present, production of pair of Higgs bosons in EFT
is known to  N$^3$LO level \cite{Chen:2019lzz}, for NLO, NNLO, see 
\cite{deFlorian:2013uza,Grigo:2015dia,deFlorian:2013jea, 
Grigo:2013rya,Frederix:2014hta,Maltoni:2014eza,Degrassi:2016vss, Borowka:2016ehy, Borowka:2016ypz}.  
All the two loop virtual amplitudes for $g+g\to hh$ that are required for the N$^3$LO cross section 
for the di-Higgs production were obtained in \cite{Banerjee:2018lfq}.
The production of di-Higgs bosons through bottom quark annihilation was 
obtained up to NNLO level in \cite{H:2018hqz}. 
In \cite{Li:2013flc,Maierhofer:2013sha,deFlorian:2015moa}, the fully differential results at NNLO level 
are presented.  While, there have been flurry of activities in the context of scalar Higgs boson, 
very little is known for the production of pair of pseudo scalar Higgs bosons at the LHC so far. 
In \cite{Dawson:1998py}, LO contribution keeping finite top mass  and NLO contributions  
using EFT framework where top quark degrees of freedom are integrated out 
have been obtained.  Like the production of single pseudo scalar Higgs boson, pair production
is also important to understand the nature of the extended Higgs sector.  In order to reduce 
the theoretical uncertainties, it is important to have QCD radiative corrections under control.  
Due to EFT, it is now possible to go beyond NLO with available
tools to make precise as well as stable predictions with respect to the unphysical scales.  
At NNLO level, we require two loop virtual, one loop single real emission and double real emission amplitudes.
In this article, as a first step towards obtaining going beyond NLO QCD corrections, we
compute all the one and two loop amplitudes that can contribute to the pure virtual part of the
cross section in dimensional regularisation and perform ultraviolet (UV) renormalisation to obtain UV finite results.

The paper is organised as follows:  in section-2, we describe how two loop virtual amplitudes
are computed. In particular, we introduce the effective Lagrangian, the relevant kinematics,
describe how projector method can be applied to obtain the scalar parts of the amplitudes, the
subtleties involved in defining the Levi-Civita tensor and $\gamma_5$ in dimensional
regularisation, ultraviolet renormalisation of strong coupling, over all
renormalisations for the composite operators and finite renormalisation for the $\gamma_5$.
In section-3, computation of the amplitudes and their infrared (IR) structure are 
briefly discussed.  In section-4, we summarise our results and conclude.

\section{Theoretical framework}
\label{sec:framework}

\subsection{Effective Lagrangian}

We work with the effective Lagrangian 
\cite{Chetyrkin:1998mw} that describes the interaction of the 
pseudo-scalar field $\Phi^A (x)$ with the gauge field $G^{a \mu\nu}$ 
and the fermion $\psi$:
\begin{eqnarray}
\label{eq:EL}
        {\cal L}^A_{eff}  = \Phi^A (x) \left[ 
-\frac{1}{8} C_G O_G (x) - \frac{1}{2} C_J O_J (x) \right] \,.
\end{eqnarray}
The pseudo-scalar gluonic ($O_G (x)$) and the light quark 
($O_J (x)$) operators are defined as
\begin{equation}
O_G(x)=G^{a\mu\nu}\tilde{G}_{\mu\nu}^a=\epsilon_{\mu\nu\rho\sigma}
G^{a\mu\nu}G^{a\rho\sigma},
\quad 
G^{a\mu\nu}=\partial^{\mu}G^{a\nu}-\partial^{\nu}G^{a\mu}+g_s f^{abc}
G_{b}^{\mu}G_{c}^{\nu},
\end{equation}
where $f^{abc}$ is the SU(3) structure constant and 
$\epsilon_{\mu\nu\rho\sigma}$ is the Levi-Civita tensor.  The
pseudo-scalar fermionic operator is the derivative of the flavour singlet
axial vector current
\begin{equation}
O_{J}(x)=\partial_{\mu}\left(\bar{\psi}\gamma^{\mu}\gamma^{5}\psi\right) \,.
\label{eq:Fields EL}
\end{equation}
The effective Lagrangian is obtained after
integrating out the top quark fields in the large top mass limit.  
Hence, the corresponding Wilson coefficients
C$_{G}$ and C$_{J}$ depend on the mass of the top quark $m_t$.
As a result of the Adler-Bardeen theorem \cite{Adler:1969gk}, 
there is no QCD correction to C$_{G}$ beyond one-loop level. On the
other hand, C$_{J}$ begins only at second-order in the strong coupling
constant $a_{s}\equiv g_{s}^{2}/16\pi^{2}=\alpha_{s}/4\pi$. The Wilsons coefficients
are given by
\begin{align}
C_{G}\left(a_{s}\right) & =-a_{s}2^{\tfrac{5}{4}}G_{F}^{\tfrac{1}{2}}\cot\beta \,,
\label{WCG}
\\
C_{J}\left(a_{s}\right) & =-\left[a_{s}C_{F}\left(\dfrac{3}{2}-3\ln\dfrac{\mu_{R}^{2}}{m_{t}^{2}}\right)
+a_{s}^{2}C_{J}^{\left(2\right)}+...\right]C_{G}\,,
\label{WCJ}
\end{align}
where $G_F$ is the Fermi constant, $\cot \beta$ -- the ratio 
of the vacuum expectation values of the two Higgs doublets, in a model where the CP
is not spontaneously broken.
$C_F$ is the quadratic Casimir in the fundamental representation
of QCD and $\mu_R$ is the renormalisation scale at which $a_{s}$ is
renormalised.

We use the effective lagrangian \ref{eq:EL} to obtain amplitudes for the production
of pair of pseudo-scalar Higgs bosons $A$ of mass $m_A$ up to two loop level in perturbative QCD. 
We restrict ourselves to the dominant gluon fusion subprocess: 
\begin{eqnarray}
\label{eq:amp}
	g(p_1)+g(p_2) \rightarrow A(p_3)+A(p_4)\,,
\end{eqnarray}
where $p_1$ and $p_2$ are the momenta of the incoming gluons, $p_{1,2}^2=0$ and $p_3$
and $p_4$ are the momenta of the outgoing pseudo-scalar Higgs bosons, $p_{3,4}^2=m_A^2$.
The Mandelstam variables for the above process are given
by
\begin{eqnarray}
s = (p_1+p_2)^2, \quad t = (p_1-p_3)^2, \quad u=(p_2-p_3)^2\,,
\end{eqnarray}
which satisfy $s+t+u=2 m_A^2$.
It is convenient to express these amplitudes in terms of the
dimensionless variables $x$, $y$ and $z$ as
\begin{eqnarray}
	s = m_A^2 {(1+x)^2 \over x},\quad t = -m_A^2 y,\quad u= -m_A^2 z \,,
\end{eqnarray}
which lead to the constraint $x^{-1} + x = y +z$.

As in the case of di-Higgs production amplitude {\it via} gluon fusion 
\cite{Glover:1987nx}, the di-pseudo scalar production amplitude,
can also be decomposed in terms of two second rank Lorentz
tensors ${\cal T}_i^{\mu \nu}$ ($i=1,2$), as follows:
\begin{eqnarray}
	{\cal M}^{\mu \nu}_{ab} \epsilon_\mu (p_1) \epsilon_\nu (p_2)= \delta_{ab} \left({\cal T}_1^{\mu\nu}~{\cal M}_1  
			    + {\cal T}_2^{\mu\nu}~{\cal M}_2\right) 
\epsilon_\mu (p_1) \epsilon_\nu (p_2)\,, 
\end{eqnarray}
where $\epsilon_\mu (p_i)$ are the polarisation vectors of the initial state
gluons.  The Lorentz scalar functions ${\cal M}_i$, $i=1,2$ are independently gauge invariant.
$\delta_{ab}$ indicates that there is no colour flow from initial
to final state.  The second rank tensors are given by
\begin{eqnarray}
\!\!\!\! {\cal T}_1^{\mu\nu} & \!=\! & g^{\mu \nu} - { p_1^\nu p_2^\mu 
                                \over p_1\cdot p_2}\,,
	\\
	\! {\cal T}_2^{\mu\nu} & \!=\! & g^{\mu \nu} + {1 \over p_1\cdot p_2~ p_T^2} \Big(
	m_A^2~ p_2^\mu p_1^\nu - 2 p_1 \cdot p_3~ p_2^\mu p_3^\nu -2 p_2\cdot p_3~ p_3^\mu p_1^\nu
	+2 p_1\cdot p_2~ p_3^\mu p_3^\nu \Big)\,,
\nonumber
\\
\end{eqnarray}
with $p_T^2 = (t u - m_A^4)/s$ is the transverse momentum square of the 
pseudo-scalar Higgs boson expressed in terms of the Mandelstam variables.
The tensor ${\cal T}_1^{\mu\nu}$ depends only on the initial state momenta $p_{1,2}$.
Using momentum conservation, it can be seen that ${\cal T}_2^{\mu\nu}$
is symmetric under the interchange of the two pseudo-scalar Higgs momenta.
The scalar functions ${\cal M}_{1,2}$ can be obtained from 
${\cal M}^{\mu \nu}_{ab}$, by using appropriate $d$-dimensional projectors 
$P_{i,ab}^{\mu \nu}$ with $i=1,2$, respectively and the projectors are given by:
\begin{eqnarray}
	P_{1,ab}^{\mu \nu} &=& \frac{\delta_{ab}}{N^2-1}
                        \left(
\frac{1}{4}\frac{d-2}{d-3} {\cal T}_1^{\mu\nu} -\frac{1}{4}\frac{d-4}{d-3}{\cal T}_2^{\mu\nu}
                        \right) \,,
	\nonumber \\
	P_{2,ab}^{\mu \nu} &=& \frac{\delta_{ab}}{N^2-1}
                        \left(
-\frac{1}{4}\frac{d-4}{d-3}{\cal T}_1^{\mu\nu}+\frac{1}{4}\frac{d-2}{d-3}{\cal T}_2^{\mu\nu}
                        \right) \,,
\label{eq:proj}
\end{eqnarray}
where $N$ corresponds to the $SU(N)$ colour group.

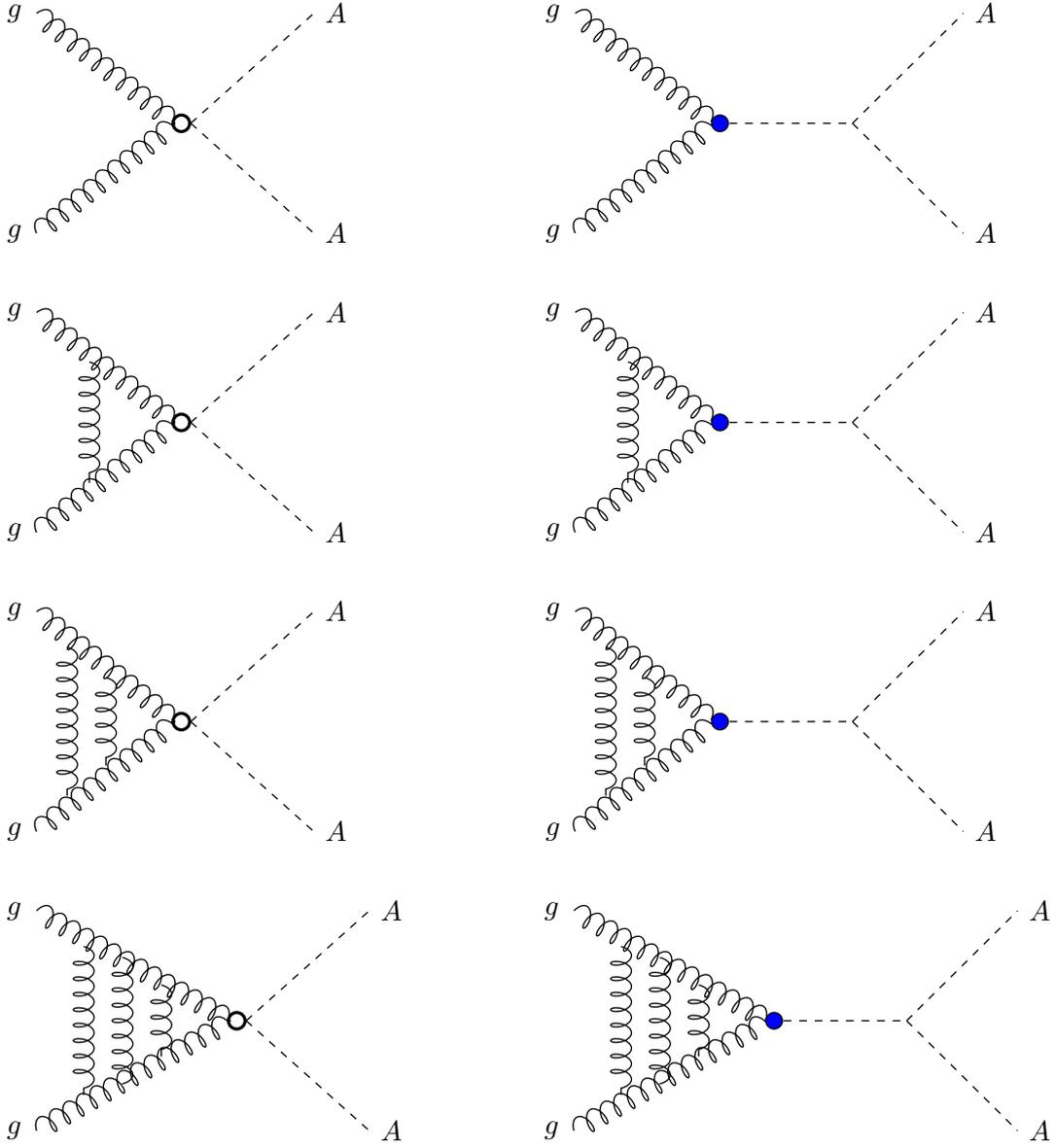
\begin{figure}[htb!]
\begin{centering}
\begin{tikzpicture}[line width=0.5 pt, scale=0.75]
\draw[gluon] (-2.5,2.0) -- (0,0);
\draw[gluon] (-2.5,-2.0) -- (0,0);
\draw[very thick] (0.11,0) circle (.15cm);
\draw[scalarnoarrow] (0.28,0)--(2.5,2.0);
\draw[scalarnoarrow] (0.28,0)--(2.5,-2.0);
\node at (-2.9,2.0) {$g$};
\node at (-2.9,-2.0) {$g$};
\node at (2.9,2.0) {$A$};
\node at (2.9,-2.0) {$A$};
 \end{tikzpicture}
\quad \quad \qquad \qquad 
\begin{tikzpicture}[line width=0.5 pt, scale=0.75]
\draw[gluon] (-2.5,2.0) -- (0,0);
\draw[gluon] (-2.5,-2.0) -- (0,0);
\draw[fill=blue] (0.11,0) circle (.15cm);
\draw[scalarnoarrow] (0.28,0)--(2.5,0);
\draw[scalarnoarrow] (2.5,0)--(4.5,2.0);
\draw[scalarnoarrow] (2.5,0)--(4.5,-2.0);
\node at (-2.9,2.0) {$g$};
\node at (-2.9,-2.0) {$g$};
\node at (4.9,2.0) {$A$};
\node at (4.9,-2.0) {$A$};
 \end{tikzpicture}
 \end{centering}
%
%
\\
\\
%
%
\begin{centering}
\begin{tikzpicture}[line width=0.5 pt, scale=0.75]
\draw[gluon] (-2.5,2.0) -- (0,0);
\draw[gluon] (-2.5,-2.0) -- (0,0);
\draw[gluon] (-1.55,1.1) -- (-1.55,-1.1);
\draw[very thick] (0.11,0) circle (.15cm);
\draw[scalarnoarrow] (0.28,0)--(2.5,2.0);
\draw[scalarnoarrow] (0.28,0)--(2.5,-2.0);
\node at (-2.9,2.0) {$g$};
\node at (-2.9,-2.0) {$g$};
\node at (2.9,2.0) {$A$};
\node at (2.9,-2.0) {$A$};
 \end{tikzpicture}
\quad \quad \qquad \qquad 
\begin{tikzpicture}[line width=0.5 pt, scale=0.75]
\draw[gluon] (-2.5,2.0) -- (0,0);
\draw[gluon] (-2.5,-2.0) -- (0,0);
\draw[gluon] (-1.55,1.1) -- (-1.55,-1.1);
\draw[fill=blue] (0.11,0) circle (.15cm);
\draw[scalarnoarrow] (0.28,0)--(2.5,0);
\draw[scalarnoarrow] (2.5,0)--(4.5,2.0);
\draw[scalarnoarrow] (2.5,0)--(4.5,-2.0);
\node at (-2.9,2.0) {$g$};
\node at (-2.9,-2.0) {$g$};
\node at (4.9,2.0) {$A$};
\node at (4.9,-2.0) {$A$};
 \end{tikzpicture}
 \end{centering}
\\
\\
%
\begin{centering}
\begin{tikzpicture}[line width=0.5 pt, scale=0.75]
\draw[gluon] (-2.5,2.0) -- (0,0);
\draw[gluon] (-2.5,-2.0) -- (0,0);
\draw[gluon] (-1.25,0.80) -- (-1.25,-0.80);
\draw[gluon] (-1.95,1.35) -- (-1.95,-1.35);
\draw[very thick] (0.11,0) circle (.15cm);
\draw[scalarnoarrow] (0.28,0)--(2.5,2.0);
\draw[scalarnoarrow] (0.28,0)--(2.5,-2.0);
\node at (-2.9,2.0) {$g$};
\node at (-2.9,-2.0) {$g$};
\node at (2.9,2.0) {$A$};
\node at (2.9,-2.0) {$A$};
 \end{tikzpicture}
\quad \quad  \qquad \qquad 
\begin{tikzpicture}[line width=0.5 pt, scale=0.75]
\draw[gluon] (-2.5,2.0) -- (0,0);
\draw[gluon] (-2.5,-2.0) -- (0,0);
\draw[gluon] (-1.25,0.80) -- (-1.25,-0.80);
\draw[gluon] (-1.95,1.35) -- (-1.95,-1.35);
\draw[fill=blue] (0.11,0) circle (.15cm);
\draw[scalarnoarrow] (0.28,0)--(2.5,0);
\draw[scalarnoarrow] (2.5,0)--(4.5,2.0);
\draw[scalarnoarrow] (2.5,0)--(4.5,-2.0);
\node at (-2.9,2.0) {$g$};
\node at (-2.9,-2.0) {$g$};
\node at (4.9,2.0) {$A$};
\node at (4.9,-2.0) {$A$};
 \end{tikzpicture}
 \end{centering}
%
%
\\
\\
%
\begin{centering}
\begin{tikzpicture}[line width=0.5 pt, scale=0.75]
\draw[gluon] (-3.5,2.0) -- (0,0);
\draw[gluon] (-3.5,-2.0) -- (0,0);
\draw[gluon] (-1.25,0.65) -- (-1.25,-0.65);
\draw[gluon] (-1.95,1.15) -- (-1.95,-1.15);
\draw[gluon] (-2.65,1.35) -- (-2.65,-1.35);
\draw[very thick] (0.11,0) circle (.15cm);
\draw[scalarnoarrow] (0.28,0)--(2.5,2.0);
\draw[scalarnoarrow] (0.28,0)--(2.5,-2.0);
\node at (-3.9,2.0) {$g$};
\node at (-3.9,-2.0) {$g$};
\node at (2.9,2.0) {$A$};
\node at (2.9,-2.0) {$A$};
 \end{tikzpicture}
\quad \quad \qquad
\begin{tikzpicture}[line width=0.5 pt, scale=0.75]
\draw[gluon] (-3.5,2.0) -- (0,0);
\draw[gluon] (-3.5,-2.0) -- (0,0);
\draw[gluon] (-1.25,0.65) -- (-1.25,-0.65);
\draw[gluon] (-1.95,1.15) -- (-1.95,-1.15);
\draw[gluon] (-2.65,1.35) -- (-2.65,-1.35);
\draw[fill=blue] (0.11,0) circle (.15cm);
\draw[scalarnoarrow] (0.28,0)--(2.5,0);
\draw[scalarnoarrow] (2.5,0)--(4.5,2.0);
\draw[scalarnoarrow] (2.5,0)--(4.5,-2.0);
\node at (-3.9,2.0) {$g$};
\node at (-3.9,-2.0) {$g$};
\node at (4.9,2.0) {$A$};
\node at (4.9,-2.0) {$A$};
\end{tikzpicture}
\end{centering}
\caption{Type-Ia (left panel) corresponds to the  $AAgg$ effective vertex 
(denoted by circle) which are form factor up to 3-loop and the right panel
(Type-Ib) is related to the effective vertex $Agg$ (denoted by shaded circle)
form factor to 3-loop order.}
\label{TypeI}
\end{figure}

In the following, we briefly discuss on the type of Feynman diagrams
that contribute up to order ${\cal O} (a_s^4)$ in QCD. 
To evaluate the 4-point amplitude $g + g \to A +A $ to any order in $a_s$,
one needs to calculate the contributing diagrams to that 
particular order and evaluate the scalar functions ${\cal M}_{1,2}$,
using the projectors $P_{i,ab}^{\mu \nu}$, $i=1,2$.  
Using the effective Lagrangian eq.\~(\ref{eq:EL}), the higher order corrections to 
$g + g \to A +A $ amplitude are calculated in massless QCD.  There
are two types of diagrams that contribute to this process.  We classify them as type-I and type-II.
The form factor type diagrams where a pair of gluons annihilate to a single A, which branches
into a pair of As belong to type-I and type-II contains t and u channel diagrams where each 
A is coupled to pair of gluons, or to quarks.  
In type-I, we have two classes of diagrams: type-Ia (fig.~\ref{TypeI} left panel) which contains
only four point $AAgg$ effective vertex and type-Ib (fig.~\ref{TypeI} right panel) containing both 
$AAg$ and $AAA$ vertices.
These diagrams contribute at tree level (${\cal O} (a_s)$)
and we need to calculate them to ${\cal O} (a_s^4)$ {\em i.e.}, up to 3-loop order.  
Since these diagrams are related to form factors of $O_G$ between gluons states and $O_J$ 
between quark and gluon states, we can readily obtain them from 
\cite{Baikov:2009bg,Gehrmann:2010ue,Ahmed:2015qpa}. 

The type-II diagrams consist of (a) two $Agg$ effective vertex (fig.~\ref{TypeIIa})
and (b) one $Agg$ effective vertex and one $Aq \bar q$ effective vertex as shown
in fig.~\ref{TypeIIb}.  Due to the axial anomaly, the pseudo-scalar operator for
the gluonic field strength mixes with the divergence of the singlet axial vector
current.  The $Agg$ effective vertex is proportional to the $C_G$ Wilson coefficient
(eq.~\ref{WCG}) which is constrained to order ${\cal O} (a_s)$ due to the Alder-Bardeen
Theorem.  The tree level diagram in type-IIa (fig.\ \ref{TypeIIa}) starts at order
${\cal O} (a_s^2)$ and each higher loop order adds an order ${\cal O} (a_s)$.
The $A q \bar q$ effective vertex is proportional to $C_J$, the Wilson
coefficient (eq.~\ref{WCJ}) which starts at order ${\cal O} (a_s^2)$.  The type-IIb 
diagrams (fig.~\ref{TypeIIb}) which consist of one $Agg$ effective vertex and one 
$Aq \bar q$ effective vertex start at one loop level at ${\cal O} (a_s^4)$.

Since, type-I diagrams are known to required order in $a_s$, 
the results presented in this paper will mainly include the type-II amplitudes up to
two loops in massless perturbative QCD {\em i.e.}\ order ${\cal O} (a_s^4)$. We use 
dimensional regularisation ($d=4+\epsilon$) to regularise both UV and IR singularities
which appear as poles in $\epsilon$ in the UV, soft and collinear regions.  Since we
will have to deal with the Levi-Civita 
tensor in $O_G$ operator and $\gamma_5$ in $O_J$ operator, both of which are constructs 
inherently in 4-dimensions, a consistent method to deal with them in $4+\epsilon$ 
dimensions is essential.  We discuss the details of a consistent and practical prescription
to go over to $4+\epsilon$ and its implications in the next section.
Hence, the scalar amplitudes ${\cal M}_i$ can be written as a sum of amplitudes     
resulting from types-I and II diagrams as
\begin{eqnarray}
        {\cal M}_i = {\cal M}_i^{\rm I} + {\cal M}_i^{\rm {II}},\quad \quad \quad i=1,2\,
\end{eqnarray}
and in the following we concentrate only on ${\cal M}_i^{\rm{II}}$.

%


\begin{figure}[htb!]
\begin{centering}
\begin{tikzpicture}[line width=0.5 pt, scale=0.75]
\draw[gluon] (-2.5,1.6) -- (0,1.6);
\draw[gluon] (-2.5,-1.6) -- (0,-1.6);
\draw[gluon] (0,1.6) -- (0,-1.6);
\draw[fill=blue] (0,1.6) circle (.15cm);
\draw[fill=blue] (0,-1.6) circle (.15cm);
\draw[scalarnoarrow] (0,1.6)--(2.5,1.6);
\draw[scalarnoarrow] (0,-1.6)--(2.5,-1.6);
\node at (-2.9,1.6) {$g$};
\node at (-2.9,-1.6) {$g$};
\node at (2.9,1.6) {$A$};
\node at (2.9,-1.6) {$A$};
 \end{tikzpicture}
\quad \quad \qquad \qquad 
\begin{tikzpicture}[line width=0.5 pt, scale=0.75]
\draw[gluon] (-2.5,1.6) -- (0,1.6);
\draw[gluon] (-2.5,-1.6) -- (0,-1.6);
\draw[gluon] (0,1.6) -- (0,-1.6);
\draw[fill=blue] (0,1.6) circle (.15cm);
\draw[fill=blue] (0,-1.6) circle (.15cm);
\draw[scalarnoarrow] (0,1.6)--(2.5,-1.6);
\draw[scalarnoarrow] (0,-1.6)--(2.5,1.6);
\node at (-2.9,1.6) {$g$};
\node at (-2.9,-1.6) {$g$};
\node at (2.9,1.6) {$A$};
\node at (2.9,-1.6) {$A$};
\end{tikzpicture}
\end{centering}
\\
\\
%
\begin{centering}
\begin{tikzpicture}[line width=0.5 pt, scale=0.75]
\draw[gluon] (-2.5,1.6) -- (0,1.6);
\draw[gluon] (-2.5,-1.6) -- (0,-1.6);
\draw[gluon] (0,1.6) -- (0,-1.6);
\draw[gluon] (-1.40,1.6) -- (-1.40,-1.6);
\draw[fill=blue] (0,1.6) circle (.15cm);
\draw[fill=blue] (0,-1.6) circle (.15cm);
\draw[scalarnoarrow] (0,1.6)--(2.5,1.6);
\draw[scalarnoarrow] (0,-1.6)--(2.5,-1.6);
\node at (-2.9,1.6) {$g$};
\node at (-2.9,-1.6) {$g$};
\node at (2.9,1.6) {$A$};
\node at (2.9,-1.6) {$A$};
 \end{tikzpicture}
\quad \quad  \qquad \qquad 
\begin{tikzpicture}[line width=0.5 pt, scale=0.75]
\draw[gluon] (-2.5,1.6) -- (0,1.6);
\draw[gluon] (-2.5,-1.6) -- (0,-1.6);
\draw[gluon] (0,1.6) -- (0,-1.6);
\draw[gluon] (-1.40,1.6) -- (-1.40,-1.6);
\draw[fill=blue] (0,1.6) circle (.15cm);
\draw[fill=blue] (0,-1.6) circle (.15cm);
\draw[scalarnoarrow] (0,1.6)--(2.5,-1.6);
\draw[scalarnoarrow] (0,-1.6)--(2.5,1.6);
\node at (-2.9,1.6) {$g$};
\node at (-2.9,-1.6) {$g$};
\node at (2.9,1.6) {$A$};
\node at (2.9,-1.6) {$A$};
\end{tikzpicture}
\end{centering}
\\
\\
%
%
\begin{centering}
\begin{tikzpicture}[line width=0.5 pt, scale=0.75]
\draw[gluon] (-2.5,1.6) -- (0,1.6);
\draw[gluon] (-2.5,-1.6) -- (0,-1.6);
\draw[gluon] (0,1.6) -- (0,-1.6);
\draw[gluon] (-0.9,1.6) -- (-0.9,-1.6);
\draw[gluon] (-1.8,1.6) -- (-1.8,-1.6);
\draw[fill=blue] (0,1.6) circle (.15cm);
\draw[fill=blue] (0,-1.6) circle (.15cm);
\draw[scalarnoarrow] (0,1.6)--(2.5,1.6);
\draw[scalarnoarrow] (0,-1.6)--(2.5,-1.6);
\node at (-2.9,1.6) {$g$};
\node at (-2.9,-1.6) {$g$};
\node at (2.9,1.6) {$A$};
\node at (2.9,-1.6) {$A$};
 \end{tikzpicture}
\quad \quad \qquad \qquad 
\begin{tikzpicture}[line width=0.5 pt, scale=0.75]
\draw[gluon] (-2.5,1.6) -- (0,1.6);
\draw[gluon] (-2.5,-1.6) -- (0,-1.6);
\draw[gluon] (0,1.6) -- (0,-1.6);
\draw[gluon] (-0.9,1.6) -- (-0.9,-1.6);
\draw[gluon] (-1.8,1.6) -- (-1.8,-1.6);
\draw[fill=blue] (0,1.6) circle (.15cm);
\draw[fill=blue] (0,-1.6) circle (.15cm);
\draw[scalarnoarrow] (0,1.6)--(2.5,-1.6);
\draw[scalarnoarrow] (0,-1.6)--(2.5,1.6);
\node at (-2.9,1.6) {$g$};
\node at (-2.9,-1.6) {$g$};
\node at (2.9,1.6) {$A$};
\node at (2.9,-1.6) {$A$};
\end{tikzpicture}
\end{centering}
\caption{Type-IIa: Sample diagrams of amplitudes up to two-loop involving two
$Agg$ effective vertex.}
\label{TypeIIa}
\end{figure}
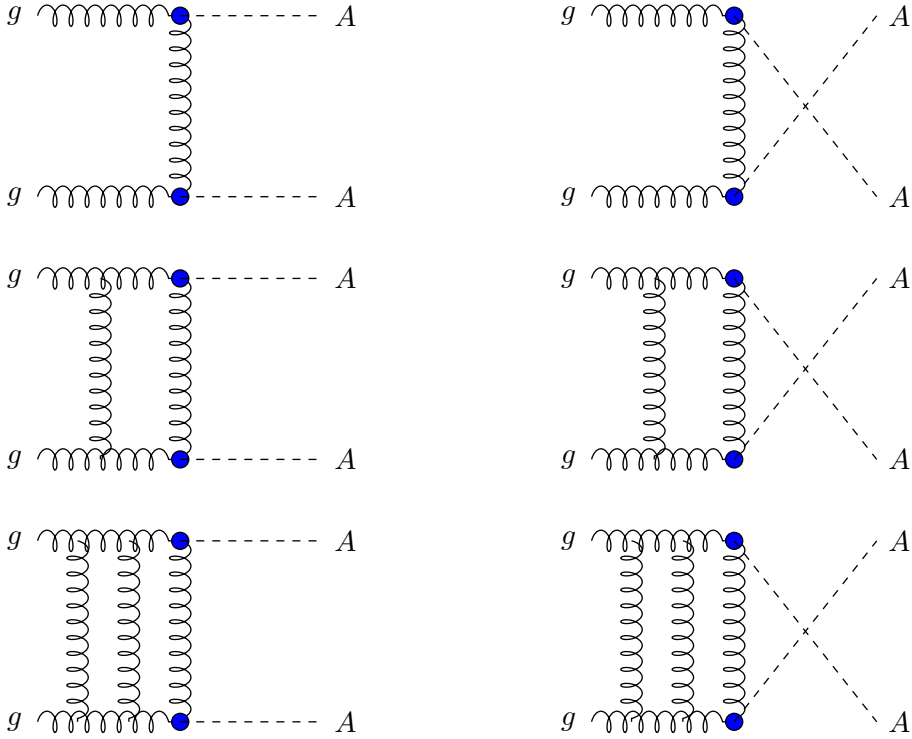
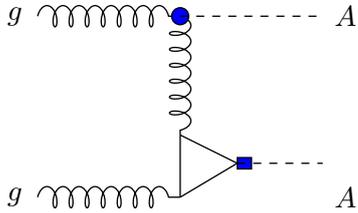
\begin{figure}[htb!]
\begin{centering}
\begin{tikzpicture}[line width=0.5 pt, scale=0.75]
\draw[gluon] (-2.5,1.6) -- (0,1.6);
\draw[gluon] (-2.5,-1.6) -- (0,-1.6);
\draw[gluon] (0,1.6) -- (0,-0.5);
\draw[fermionnoarrow] (0,-0.5) -- (0,-1.6);
\draw[fermionnoarrow] (0,-0.5) -- (1,-1);
\draw[fermionnoarrow] (1,-1) -- (0,-1.6);
\draw[fill=blue] (0,1.6) circle (.15cm);
\draw[fill=blue] (1,-0.9) rectangle (1.25,-1.1);
\draw[scalarnoarrow] (0,1.6)--(2.5,1.6);
\draw[scalarnoarrow] (1,-1)--(2.5,-1);
\node at (-2.9,1.6) {$g$};
\node at (-2.9,-1.6) {$g$};
\node at (2.9,1.6) {$A$};
\node at (2.9,-1.6) {$A$};
 \end{tikzpicture}
\end{centering}
\caption{Type-IIb Diagram involving mixing of the effective vertices $Agg$ and $A q \bar q$
(denoted by shaded rectangle) which contribute at ${\cal O} (a_s^4)$}
\label{TypeIIb}
\end{figure}

\subsection{$\gamma_5$ within dimensional regularisation}

Due to the axial anomaly, the pseudo-scalar gluonic operator 
$O_G=\epsilon_{\mu\nu\rho\sigma} G^{a\mu\nu} G^{a\rho\sigma} $
is related to the divergence of the
axial vector current $O_J= \partial_\mu(\bar \psi \gamma^\mu \gamma_5 \psi)$.  
Computation of 
higher order corrections with chiral quantities, involve inherently 
$d=4$ dimensional objects like $\gamma_5$ and the Levi-Civita tensor
$\epsilon_{\mu\nu\rho\sigma}$,
and this warrants a prescription in going away from 4-dimension 
{\it i.e.}~$d= 4 +\epsilon$.
There exist several prescriptions to deal with $\gamma_5$ in dimensional
regularisation \cite{tHooft:1972tcz,Korner:1991sx}.
In multi-loop computations that use dimensional regularisation, 
we use the self-consistent 
 prescription for $\gamma_5$ that was proposed by 't~Hooft and Veltman
\cite{tHooft:1972tcz}.  In this prescription, one defines $\gamma_5$ as
\begin{eqnarray} 
\gamma_5=\frac{i}{4!} \epsilon_{\mu_1\mu_2\mu_3\mu_4} 
\gamma^{\mu_1}
\gamma^{\mu_2}
\gamma^{\mu_3}
\gamma^{\mu_4} \,,
 \label{g5}
\end{eqnarray} 
where Levi-Civita tensor is purely 4-dimensional, while the Lorentz
indices on the $\gamma^{\mu_i}$ are in $d=4+\epsilon$ dimensions. To
maintain the anti-commuting nature of $\gamma_5$ with $d$-dimensional 
$\gamma^{\mu_i}$, the symmetrical form of the axial current has to
be used
\begin{eqnarray} 
J^5_\mu = 
\frac{1}{2}
\bar \psi (\gamma_\mu \gamma_5 - \gamma_5 \gamma_\mu) \psi \,,
\end{eqnarray}
this is in concurrence with the above definition of $\gamma_5$ in eq.~\ref{g5},
and will lead to
\begin{eqnarray} 
J^5_\mu = 
\frac{i}{3!} \epsilon_{\mu \nu_1\nu_2\nu_3} 
\bar \psi \gamma^{\nu_1} \gamma^{\nu_2} \gamma^{\nu_3} \psi \,.
\end{eqnarray}
The $O_G$ and  $O_J$ operators now take the form
\begin{equation}
O_G(x)=G^{a\mu\nu}\tilde{G}_{\mu\nu}^a=\epsilon_{\mu\nu\rho\sigma}
G^{a\mu\nu}G^{a\rho\sigma},
\qquad
O_{J}(x)=\frac{i}{3!} \epsilon_{\mu \nu_1\nu_2\nu_3} 
\partial^{\mu}\left(\bar{\psi}  \gamma^{\nu_1} \gamma^{\nu_2}
 \gamma^{\nu_3} \psi\right) \,.
\label{eq:Fields EL}
\end{equation}
Contraction of two Levi-Civita tensors that result from either $O_G$
operator or the mixing of $O_G$ and $O_J$ operators is given
by
\noindent
\begin{equation}
\epsilon_{\mu_{1}\nu_{1}\rho_{1}\sigma_{1}}\epsilon^{\mu_{2}\nu_{2}\rho_{2}\sigma_{2}}=\left|\begin{array}{cccc}
\delta_{\mu_{1}}^{\mu_{2}} & \delta_{\mu_{1}}^{\nu_{2}} & \delta_{\mu_{1}}^{\rho_{2}} & \delta_{\mu_{1}}^{\sigma_{2}}\\
\delta_{\nu_{1}}^{\mu_{2}} & \delta_{\nu_{1}}^{\nu_{2}} & \delta_{\nu_{1}}^{\rho_{2}} & \delta_{\nu_{1}}^{\sigma_{2}}\\
\delta_{\rho_{1}}^{\mu_{2}} & \delta_{\rho_{1}}^{\nu_{2}} & \delta_{\rho_{1}}^{\rho_{2}} & \delta_{\rho_{1}}^{\sigma_{2}}\\
\delta_{\sigma_{1}}^{\mu_{2}} & \delta_{\sigma_{1}}^{\nu_{2}} & \delta_{\sigma_{1}}^{\rho_{2}} & \delta_{\sigma_{1}}^{\sigma_{2}}
\end{array}\right|\,,
\end{equation}
the Lorentz indices in this determinant, could now be considered as
$d$-dimensional and the consequence would be, addition of only the
inessential ${\cal O} (\epsilon)$ terms to the renormalisated quantity
\cite{Larin:1993tq}.  This 
prescription though is not without consequence-- a finite
renormalisation of the axial vector current \cite{Larin:1991tj}
is required in order to fulfill the chiral Ward identities and the
Adler-Bardeen theorem.  This will be discussed further in the next
section.

\subsection{UV renormalisation, operator renormalization and 
mixing}
\label{sec:ren}
In dimensional regularisation with $d=4+\epsilon$, the bare strong
coupling constant denoted by $\hat a_s$ is related to its renormalized
coupling by $a_s$
\begin{align}
  \label{eq:asAasc}
  {\hat a}_{s} S_{\epsilon} = \left( \frac{\mu^{2}}{\mu_{R}^{2}}  \right)^{\epsilon/2}
  Z_{a_{s}} a_{s} \,,
\end{align}
with
$S_{\epsilon} = {\rm exp} \left[ (\gamma_{E} - \ln 4\pi)\epsilon/2
\right]$  with $\gamma_E \approx 0.5772...$ the Euler-Mascheroni constant
and $\mu$ is the scale introduced to keep the strong coupling constant
dimensionless in $d=4+\epsilon$ space-time dimensions.  The
renormalisation constant $Z_{a_{s}}$~\cite{Tarasov:1980au} is given by
\begin{align}
  \label{Zas}
  Z_{a_{s}}&= 1+ a_s\left[\frac{2}{\epsilon} \beta_0\right]
             + a_s^2 \left[\frac{4}{\epsilon^2 } \beta_0^2
             + \frac{1}{\epsilon}  \beta_1 \right]
             + a_s^3 \left[\frac{8}{ \epsilon^3} \beta_0^3
             +\frac{14}{3 \epsilon^2}  \beta_0 \beta_1 +  \frac{2}{3
             \epsilon}   \beta_2 \right] \,,
\end{align}
up to ${\cal O}(a_{s}^{3})$. $\beta_{i}$ are the coefficients of the
QCD $\beta$ function and are given by~\cite{Tarasov:1980au}
\begin{align}
  \beta_0&={11 \over 3 } C_A - {4 \over 3 } n_f T_{F}\, ,
           \nonumber \\[0.5ex]
  \beta_1&={34 \over 3 } C_A^2- 4 n_f C_F T_{F} -{20 \over 3} n_f
           T_{F} C_A \, ,
           \nonumber \\[0.5ex]
  \beta_2&={2857 \over 54} C_A^3 
           -{1415 \over 27} C_A^2 n_f T_{F}
           +{158\over 27} C_A n_f^2 T_{F}^{2}
               +{44 \over 9} C_F n_f^2 T_{F}^{2}
               \nonumber\\
&-{205 \over 9} C_F C_A n_f T_{F}
               + 2 C_F^2 n_f T_{F} \,,
\end{align}
where $n_f$ is the number of flavors and $T_F = 1/2$.  
As we work in an effective theory obtained after integrating out the 
top quark fields in the large top quark mass limit, $n_f=5$.
The Casimirs
of SU(N) are given by $C_F$ and $C_A$:
\begin{eqnarray}
C_F = {N^2-1\over 2 N},  \quad \quad C_A = N \, .
\end{eqnarray}
For type-I diagrams which begin to contribute at LO, the $Z_{a_s}$ up to
order ${\cal O}(a_{s}^{3})$ will be needed while for type-II 
diagrams, one order lower is sufficient.

Apart from the renormalisation of strong coupling in the massless QCD, 
the amplitudes require the renormalisation of vertices resulting from the  composite
operators $O_G$ and $O_J$ of the effective Lagrangian eq.~(\ref{eq:EL}).  
The renormalised operators are denoted by $[~]$ parenthesis, while the
bare quantities without the parenthesis.

The renormalisation of $O_J$ is related to the renormalisation of the 
singlet axial vector current $J^\mu_5$ which needs the standard overall
UV renormalisation constant $Z^s_{\overline{MS}}$ and a finite
renormalisation constant $Z^s_5$.  The later is 
necessary in dimensional
regularisation in order to ensure the nature of operator
relation resulting from  axial anomaly \cite{Adler:1969er}
\begin{align}
  \label{eq:Anomaly}
  \left[ \partial_{\mu}J^{\mu}_{5} \right] = a_{s} \frac{n_{f}}{2} \left[ G\tilde{G} \right]\,,
\qquad
\qquad
  \text{i.e.}~~~ \left[ O_{J} \right] = a_{s} \frac{n_{f}}{2} \left[ O_{G} \right]\,,
\end{align}
which is true in Pauli-Villars, a 4-dimensional regularisation. 
To preserve eq.~\ref{eq:Anomaly} in $4+\epsilon$ dimensions, the multiplicative finite renormalisation 
constant $Z^s_5$ is required.
The bare operator $O_J$ is renormalised
multiplicatively, exactly in the same way as the singlet axial
vector current $J^\mu_5$, through
\begin{align}
  \label{OJRen}
  \left[ O_J \right] = Z^s_5 ~Z^s_{\overline{MS}} ~O_J\,,
\end{align}
whereas the bare pseudo-scalar gluon operator $O_G$
mixes with fermionic operator $O_J$ under the renormalisation through
\begin{align}
  \left[ O_G \right] = Z_{GG} ~O_G + Z_{GJ} ~O_J \,,
  \label{OGRen}
\end{align}
with the corresponding renormalisation constants $Z_{GG}$ and
$Z_{GJ}$. 
Combining the above two equations in a matrix form, we have
\begin{equation}
  \label{eq:OpMat}
  \left[ O_i \right] = Z_{ij} ~O_j  \,,
\qquad \qquad {\rm where} \quad i,j = \{G, J\}\,,
\end{equation}
\begin{align}
  O \equiv
  \begin{pmatrix}
    O_G\\
    O_J
  \end{pmatrix}
  \qquad\quad &\text{and}  \qquad\quad
                Z \equiv
                \begin{pmatrix}
                  Z_{GG} & Z_{GJ}\\
                  Z_{JG} & Z_{JJ}
                \end{pmatrix}\,,
\end{align}
where $Z_{JG} = 0$ to all orders in perturbation theory and 
$Z_{JJ} \equiv Z^s_5 Z^s_{\overline{MS}}$.  The renormalisation constants
required for above equation are available up to 
${\cal O} (a_s^3)$ \cite{Larin:1993tq},
\cite{Zoller:2013ixa} which was computed using OPE. 
For earlier works on this, see \cite{Kataev:1981aw,Kataev:1981gr}. 
Using a completely 
different method the same quantities were calculated by some of us
\cite{Ahmed:2015qpa}
and found to be in full agreement. 
The UV renormalisation constant of the singlet axial vector current $J^\mu_5$ in the 
$\overline{MS}$ scheme is
\begin{align}
Z_{\overline{MS}}^s = & 1+a_{s}^{2}\left[C_{A}C_{F}\left\{
-\dfrac{44}{3\epsilon}\right\} +C_{F}n_{f}\left\{
-\dfrac{10}{3\epsilon}\right\} \right]\nonumber \\
 & +a_{s}^{3}\left[C_{A}^{2}C_{F}\left\{
-\dfrac{1936}{27\epsilon^{2}}-\dfrac{7156}{81\epsilon}\right\}
+C_{F}^{2}n_{f}\left\{ \dfrac{44}{9\epsilon}\right\} +C_{F}n_{f}^{2}\left\{
\dfrac{80}{27\epsilon^{2}}-\dfrac{52}{81\epsilon}\right\} \right]\nonumber \\
 & +a_{s}^{3}\left[C_{A}C_{F}^{2}\left\{ \dfrac{616}{9\epsilon}\right\}
+C_{A}C_{F}n_{f}\left\{
-\dfrac{88}{27\epsilon^{2}}-\dfrac{298}{81\epsilon}\right\} \right] \,,
\end{align}
and the finite renormalisation constant $Z_5^s$ is
\begin{equation}
Z_5^s=1+a_s\left\{ -4C_{F}\right\} +a_s^2 \left\{
22 C_F^2-\dfrac{107}{9}C_A C_F+\dfrac{31}{18}C_F n_f \right\} \,.
\end{equation}
The renormalisation constants for $O_G$ and $O_J$ operators up to two loops
are given by
\begin{align}
  \label{ZGGtZGJ}
  Z_{GG} &= 1 +  a_s \Bigg[ \frac{22}{3\epsilon}
           C_{A}  -
           \frac{4}{3\epsilon} n_{f} \Bigg] 
           + 
           a_s^2 \Bigg[ \frac{1}{\epsilon^2}
           \Bigg\{ \frac{484}{9} C_{A}^2 - \frac{176}{9} C_{A}
           n_{f} + \frac{16}{9} n_{f}^2 \Bigg\}
           \nonumber\\&
           + \frac{1}{\epsilon} \Bigg\{ \frac{34}{3} C_{A}^2  
         -\frac{10}{3} C_{A} n_{f}  - 2 C_{F} n_{f} \Bigg\} \Bigg],
         \nonumber\\
  Z_{GJ} &=  a_s \Bigg[ - \frac{24}{\epsilon} C_{F} \Bigg]
                    + 
                    a_s^2 \Bigg[ \frac{1}{\epsilon^2}
                    \Bigg\{ - 176 C_{A} C_{F} + 32 C_{F} n_{f} \Bigg\}
                    \nonumber\\&
                    + \frac{1}{\epsilon} \Bigg\{ - \frac{284}{3} C_{A} C_{F} +
                    84 C_{F}^2 + \frac{8}{3} C_{F} n_{f} \Bigg\}  \Bigg],
                    \nonumber\\
   Z_{JJ} &= 1 + a_s\left[-4 C_F\right] + a_s^2\Bigg[  
-\frac{44}{3\epsilon}C_A C_F - \frac{10}{3\epsilon}C_F n_f   
   \nonumber\\&
   + 22C_F^2 -\frac{107}{9}C_A C_F  + \frac{31}{18}C_F n_f \Bigg].
\end{align}
The matrix element that would contribute to the $ g + g \to A + A$ 
amplitude can be obtained {\it via} the insertion of two renormalised operators 
$[O_G]$ (eq.~\ref{OGRen}) and $[O_J]$ (eq.~\ref{OJRen}) for each
$A$, which would involve the following operator insertion between
gluon states: 
\begin{eqnarray}
\langle g|[O_G O_G] |g \rangle;  \quad
\langle g|[O_G O_J] |g \rangle \quad
{\rm and} \quad
\langle g|[O_J O_J] |g \rangle \,.
\end{eqnarray}
The above operator renormalisation constants  (eq.~\ref{ZGGtZGJ}) 
and the strong coupling renormalisation
 constant $Z_{a_s}$ (eq.~\ref{Zas}) would take care of the UV renormalisation.
The gluonic operator $O_G$ couples to gluons at LO (${\cal O} (a_s)$) and 
the fermionic operator $O_J$ couples to quarks at LO (${\cal O} (a_s^2)$).  The
basic matrix elements that have to be evaluated diagrammatically 
involve the following bare operators combination:
$\langle g|O_G^2  |g \rangle$, $\langle g|O_G O_J|g \rangle$ 
and $\langle g|O_J^2  |g \rangle$.  The $O_G^2$ starts to
contribute at tree level at ${\cal O} (a_s^2)$, $O_G O_J$
begins to contribute at one-loop level and at  ${\cal O} (a_s^4)$
while $O_J^2$ starts to contribute at ${\cal O} (a_s^6)$.  Here
we compute $g +g \to A+A$ amplitude to order ${\cal O} (a_s^4)$ and
hence the contributing terms are from $O_G^2$ calculated up to 
two-loops and the $O_G O_J$ combination from one-loop.
We need the following renormalised operator $[O_{G} O_{G} ]$
and $[ O_G O_J ]$
which is given by
\begin{eqnarray}
\label{OGOG}
\left[O_G O_G \right] &=& Z_{GG}^2 ~O_G O_G +2Z_{GG} Z_{GJ} ~O_G O_J + Z_{GJ}^2 ~O_J O_J
\,,
\nonumber 
\\
\left[O_G O_J \right] &=& Z_{GG} Z_{JJ} ~O_G O_J + Z_{GJ} Z_{JJ} ~O_J O_J\,.
\end{eqnarray}
Sandwiching $\left[O_G O_G \right]$ and $\left[O_G O_J\right]$  between gluon states and using 
eq.~(\ref{OGOG}), we obtain up to two loops: 
\noindent 
\begin{eqnarray}
\mathcal{M}^{\rm{II}}_{GG,g} & = &
Z_{GG}^{2}\Big(\hat {\mathcal{M}}_{GG,g}^{\rm{II}\left(0\right)}+\hat a_{s}\hat {\mathcal{M}}_{GG,g}^{\rm{II},\left(1\right)}
+\hat a_{s}^{2}\hat {\mathcal{M}}_{GG,g}^{\rm{II}\left(2\right)}\Big)\nonumber
\\
 &&
+2Z_{GG}Z_{GJ}\left(\hat a_{s}\hat {\mathcal{M}}_{GJ,g}^{\rm{II}\left(1\right)}+\hat a_{s}^{2}\hat {\mathcal{M}}_{GJ,g}^{\rm{II}\left(2\right)}\right)\nonumber
\\
 &&
+Z_{GJ}^{2}\Big(\hat a_{s}\hat { \mathcal{M}}_{JJ,g}^{\rm{II}\left(1\right)}
+\hat a_{s}^{2}\hat { \mathcal{M}}_{JJ,g}^{\rm{II}\left(2\right)}\Big)\,,
\nonumber\\
\mathcal{M}^{\rm{II}}_{GJ,g} & = & 
Z_{GG} Z_{JJ} \Big( \hat a_s \hat {{\cal M}}^{II(1)}_{GJ,g}
                    + \hat a_s^2 \hat {{\cal M}}^{II(2)}_{GJ,g}
              \Big)
+ Z_{GJ} Z_{JJ} \Big(\hat a_s \hat {{\cal M}}^{II(1)}_{JJ,g} + \hat a_s^2 \hat {{\cal M}}^{II(2)}_{JJ,g} \Big)\,,
\label{OGGGJ}
\end{eqnarray}
where $\mathcal{M}^{\rm{II}}_{XY,g} = \langle g|\left[O_X O_Y \right]|g\rangle$ and
$\hat {\mathcal{M}}^{\rm{II}}_{XY,g} = \langle g|O_X O_Y |g\rangle$, with $X,Y=\{G,J\}$
that contribute to the type-II diagrams.
$\hat{\mathcal{M}}_{GJ,g}^{\rm{II}\left(2\right)}$,
$\hat{\mathcal{M}}_{JJ,g}^{\rm{II}\left(1\right)}$
and
$\hat{\mathcal{M}}_{JJ,g}^{\rm{II}\left(2\right)}$
do not contribute in our case as they are of order higher than $a_{s}^{4}$ when combined with
their respective Wilson coefficients.  Finally,  $\mathcal{M}^{\rm I I}_{GG,g}$
can be expressed in powers of 
renormalised $a_s$ as
\begin{equation}
\mathcal{M}^{\rm I I}_{GG,g}=\mathcal{M}_{GG,g}^{{\rm II}\left(0\right)}+a_{s}\mathcal{M}_{GG,g}^{{\rm II}\left(1\right)}+a_{s}^{2}\mathcal{M}_{GG,g}^{{\rm II} \left(2\right)}+\mathcal{O}\left(a_{s}^{3}\right)\,.
\end{equation}
The coefficients $\mathcal{M}_{GG,g}^{{\rm II}\left(i\right)}$ can be  related to 
$\hat {\mathcal{M}}_{GG,g}^{{\rm II}\left(i\right)}$ using eq.~\ref{Zas} and eq.~\ref{ZGGtZGJ}. 
Expanding the renormalisation constants $Z_{KL}$ in eq.~\ref{ZGGtZGJ} as 
\begin{eqnarray}
Z_{KL} = \delta_{KL} + \displaystyle{\sum_{i=1}^\infty} a_s^i Z_{KL}^{(i)},
\quad \quad K,L = \{G,J\}
\end{eqnarray}
we find
\begin{eqnarray}
{\cal M}^{{\rm II}(0)}_{GG,g} &=& \hat {{\cal M}}^{{\rm II}(0)}_{GG,g}\,,
\nonumber\\
{\cal M}^{{\rm II}(1)}_{GG,g} &=& {1 \over \mu_R^\epsilon}
                           \hat {{\cal M}}^{{\rm II}(1)}_{GG,g}
                           +2 Z^{(1)}_{GG} \hat {{\cal M}}^{{\rm II}(0)}_{GG,g}\,,
\nonumber\\
{\cal M}^{{\rm II}(2)}_{GG,g} &=& {1 \over \mu_R^{2\epsilon}}
                                \hat {{\cal M}}^{{\rm II}(2)}_{GG,g}
                            +{1 \over \mu_R^{\epsilon}}
                            \Bigg( 
                             {2 \beta_0 \over \epsilon} \hat {{\cal M}}^{{\rm II}(1)}_{GG,g}
                            +2 Z^{(1)}_{GJ} \hat {{\cal M}}^{{\rm II}(1)}_{GJ,g}
                               +2 Z^{(1)}_{GG} \hat {{\cal M}}^{{\rm II}(1)}_{GG,g}\Bigg)
\nonumber\\
&&                               +\Big( 2 Z^{(2)}_{GG} + (Z^{(1)}_{GG})^2 \Big)
                                \hat {{\cal M}}^{{\rm II}(0)}_{GG,g} \,.
\end{eqnarray}
Similarly for $\mathcal{M}^{\rm I I}_{GJ,g}$,
we find
\begin{equation}
\mathcal{M}^{\rm I I}_{GJ,g}=a_{s}\mathcal{M}_{GJ,g}^{{\rm II}\left(1\right)}+a_{s}^{2}\mathcal{M}_{GJ,g}^{{\rm II} \left(2\right)}+\mathcal{O}\left(a_{s}^{3}\right) \,,
\end{equation}
where
\begin{eqnarray}
{{\cal M}}^{{\rm II}(1)}_{GJ,g}  &=& {1 \over \mu_R^\epsilon}
                                 \hat {{\cal M}}^{{\rm II}(1)}_{GJ,g}\,,
\nonumber\\
{\cal M}^{{\rm II}(2)}_{GJ,g} &=& {1 \over \mu_R^\epsilon}
                            \Bigg( {2 \beta_0 \over \epsilon} + Z^{(1)}_{JJ} + Z^{(1)}_{GG}\Bigg)
                              \hat {{\cal M}}^{{\rm II}(1)}_{GJ,g}
                           +{1 \over \mu_R^\epsilon} Z^{(1)}_{GJ} \hat {{\cal M}}^{{\rm II}(1)}_{JJ,g}
                           +{1 \over \mu_R^{2\epsilon}}
                              \hat {{\cal M}}^{{\rm II}(2)}_{GJ,g}\,.
\end{eqnarray}

We find that the UV singularities
that appear at one-loop and two-loop levels can be taken care of by
the coupling constant renormalisation $Z_{a_s}$ and operator renormalisation
$Z_{ij}$. 
At this point we would like to stress that there could be additional contact terms
required as a result of the behaviour of product of operators $O_G O_G$ or $O_G O_J$
 at short distances.  As shown in \cite{Zoller:2013ixa}, we find that there are
no contact terms as a result of these product of operators at short distances.
For earlier works on this, see \cite{Kataev:1981aw,Kataev:1981gr}. 

\section{Calculational details}

\subsection{Calculation of the Amplitude}
\label{sec:calc}

Our task of computing the amplitude $g + g \to A + A$ has reduced to the type-II diagrams up to ${\cal O} (a_s^4)$.  This involves diagrams with two $Agg$ effective vertices, up to two-loop level in QCD (Type-IIa) and diagrams with one $Agg$ effective vertex and one $Aq \bar q$ effective vertex which involves terms up to one loop in QCD (Type-IIb).  Diagrams involving two $A q \bar q$ effective vertex start
at ${\cal O}(a_s^5)$ and are not considered here.  
Applying the projectors $P_{i,ab}^{\mu \nu}$ on the amplitudes, we extract the 
scalar coefficients ${\cal M}_{i}$ with $i=1,2$ at every order in the perturbation.

All the tree level, one loop and two loop Feynman diagrams in massless QCD are generated 
using QGRAF \cite{Nogueira:1991ex} where additional vertices resulting from 
effective lagrangian eq.~(\ref{eq:EL}) are incorporated.  
There are two tree level diagrams, 35 one-loop
diagrams and 789 two-loop diagrams of type-IIa.   For type-IIb which involves
effective quark and gluon couplings to pseudo-scalar Higgs, there
are no tree level diagrams but 8 diagrams that contribute at one-loop which
suffices to generate diagrams up to ${\cal O} (a_s^4)$.  The raw QGRAF output 
is converted with the help of in-house codes based on FORM \cite{Vermaseren:2000nd} to
include appropriate Feynman rules and to perform 
trace of Dirac matrices, contraction of Lorentz
indices and colour indices.  At this stage, we encounter huge number of one and scalar 2-loop
Feynman integrals, which contain a set of propagator denominators and 
a combination of scalar products between loop momenta and independent external momenta.  
These Feynman integrals can be classified in terms
of propagator denominators, that they contain.  
It is hence important to identify the momentum shifts
that are required to express each of these diagrams in terms
of a standard set of propagators called auxiliary topology. 
We use REDUZE2 package~\cite{vonManteuffel:2012np} to achieve this.
The auxiliary topologies needed for the present case are same as those found in 
vector boson pair production \cite{Gehrmann:2013cxs,Gehrmann:2014bfa} at two loops.

As expected these large number of scalar integrals are not all independent.  To
establish the relations, some properties of the Feynman integrals in dimensional
regularisation are used.  Exploiting the fact that, the total derivative with
respect to any loop momenta of these integrals, evaluates to a surface term,
which vanishes, leads to integration-by-parts (IBP) identities \cite{Tkachov:1981wb,
Chetyrkin:1981qh}.  In addition, the fact that all integrals are Lorentz scalars,
gives rise to Lorentz invariance (LI) identities \cite{Gehrmann:1999as}.  As
a result, these integrals can in turn be expressed in terms of a much smaller set
of integrals which are irreducible and appropriately called master integrals (MI).
Several automated computer algebra packages are available \cite{Anastasiou:2004vj,
Smirnov:2008iw,Studerus:2009ye,vonManteuffel:2012np,Lee:2013mka} that use
the Laporta algorithm \cite{Laporta:2001dd} to reduce these Feynman integrals 
to the MIs.  We have used the Mathematica based package LiteRed \cite{Lee:2013mka}
to perform the reductions of all the integrals to MIs.  At one-loop, there are 10
MIs, while at two-loop the number is 149.  These two-loop MIs are the same as
two-loop four-point functions with two equal mass external legs.  The analytical
result for the each MI in terms of Laurent series expansion in $\epsilon$ is 
given in \cite{Gehrmann:2013cxs,Gehrmann:2014bfa}.

At this stage, the renormalisation of the 
strong coupling constant and of the operators $O_G$ and $O_J$, described in
section \ref{sec:ren}, removes all the UV singularities.  The singularities
that still remain are purely of infrared origin and the next section is devoted to
it.

\subsection{Infrared factorization}

The UV finite amplitudes that we have computed contain only divergences 
of infrared origin, which appear as poles in the dimensional regularization parameter $\epsilon$.  
They are expected to cancel against real emission diagrams for the IR safe observables.  
While these singularities disappear in the physical observables, the 
amplitudes beyond leading order show  a very rich universal structure in the IR. 
In \cite{Catani:1998bh}, Catani predicted the IR poles of two-loop n-point UV finite amplitudes 
in terms of certain universal IR anomalous dimensions. 
Later, in \cite{Sterman:2002qn}, factorization and resummation properties of QCD amplitudes were used
to understand the IR structure and subsequently the attempts were made to predict the structure
of IR poles beyond two loops in ~\cite{Becher:2009cu,Gardi:2009qi}. 
Following \cite{Catani:1998bh}, we obtain
\begin{eqnarray}
\label{mfin}
\mathcal{M}_i ^{\rm{II},(0)} &=& \mathcal{M}_i ^{\rm{II},(0)} \,,
\nonumber\\
\mathcal{M}_i ^{\rm{II},(1)} &=& 2\mathbf{I}_{g}^{(1)}(\epsilon)\mathcal{M}_i ^{\rm{II},(0)} + \mathcal{M}_i ^{\rm{II},(1),fin} \,,
\nonumber\\
\mathcal{M}_i ^{\rm{II},(2)} &=& 4\mathbf{I}_{g}^{(2)}(\epsilon)\mathcal{M}_{i} ^{\rm{II},(0)} + 2\mathbf{I}_{g}^{(1)}(\epsilon)\mathcal{M}_i ^{\rm{II},(1)}
+ \mathcal{M}_{i} ^{\rm{II},(2),fin} \,,
\end{eqnarray}
where $\mathbf{I}_{g}^{(1)}(\epsilon), \,\mathbf{I}_{g}^{(2)}(\epsilon) $ are the IR singularity operators given by
\begin{align}
\boldsymbol{I}_{g}^{\left(1\right)}\left(\epsilon\right) & =-\frac{e^{-\frac{\epsilon}{2}\gamma_{E}}}{\Gamma\left(1+\frac{\epsilon}{2}\right)}\left(\frac{4C_{A}}{\epsilon^{2}}-\frac{\beta_{0}}{\epsilon}\right)\left(-\frac{s}{\mu_{R}^{2}}\right)^{\frac{\epsilon}{2}} \,,\\
\boldsymbol{I}_{g}^{\left(2\right)}\left(\epsilon\right) & =-\frac{1}{2}\boldsymbol{I}_{g}^{\left(1\right)}\left(\epsilon\right)\left[\boldsymbol{I}_{g}^{\left(1\right)}\left(\epsilon\right)-\frac{2\beta_{0}}{\epsilon}\right]+\frac{e^{\frac{\epsilon}{2}\gamma_{E}}\Gamma\left(1+\epsilon\right)}{\Gamma\left(1+\frac{\epsilon}{2}\right)}\left[-\frac{\beta_{0}}{\epsilon}+K\right]\boldsymbol{I}_{g}^{\left(1\right)}\left(2\epsilon\right)+2\boldsymbol{H}_{g}^{\left(2\right)}\left(\epsilon\right) \,,\label{eq: I operators}
\end{align}
with

\noindent 
\begin{eqnarray}
K & =&\left(\frac{67}{18}-\frac{\pi^{2}}{6}\right)C_{A}-\frac{5}{9}n_{f}\,,\\
\boldsymbol{H}_{g}^{\left(2\right)}\left(\epsilon\right) & =&\left(-\frac{s}{\mu_{R}^{2}}\right)^{\epsilon}
\frac{e^{-\frac{\epsilon}{2}\gamma_{E}}}{\Gamma\left(1+\frac{\epsilon}{2}\right)}
\frac{1}{\epsilon}\left\{ C_{A}^{2}\left(-\frac{5}{24}-\frac{11}{48}\zeta_{2}
-\frac{\zeta_{3}}{4}\right)+C_{A}n_{f}\left(\frac{29}{54}+\frac{\zeta_{2}}{24}\right)
\right.
\nonumber\\
&&\left. -\frac{1}{4}C_{F}n_{f}-\frac{5}{54}n_{f}^{2}\right\} \,.
\label{eq: K and H}
\end{eqnarray}
\begin{figure}[htb!]
\includegraphics[scale=0.45]{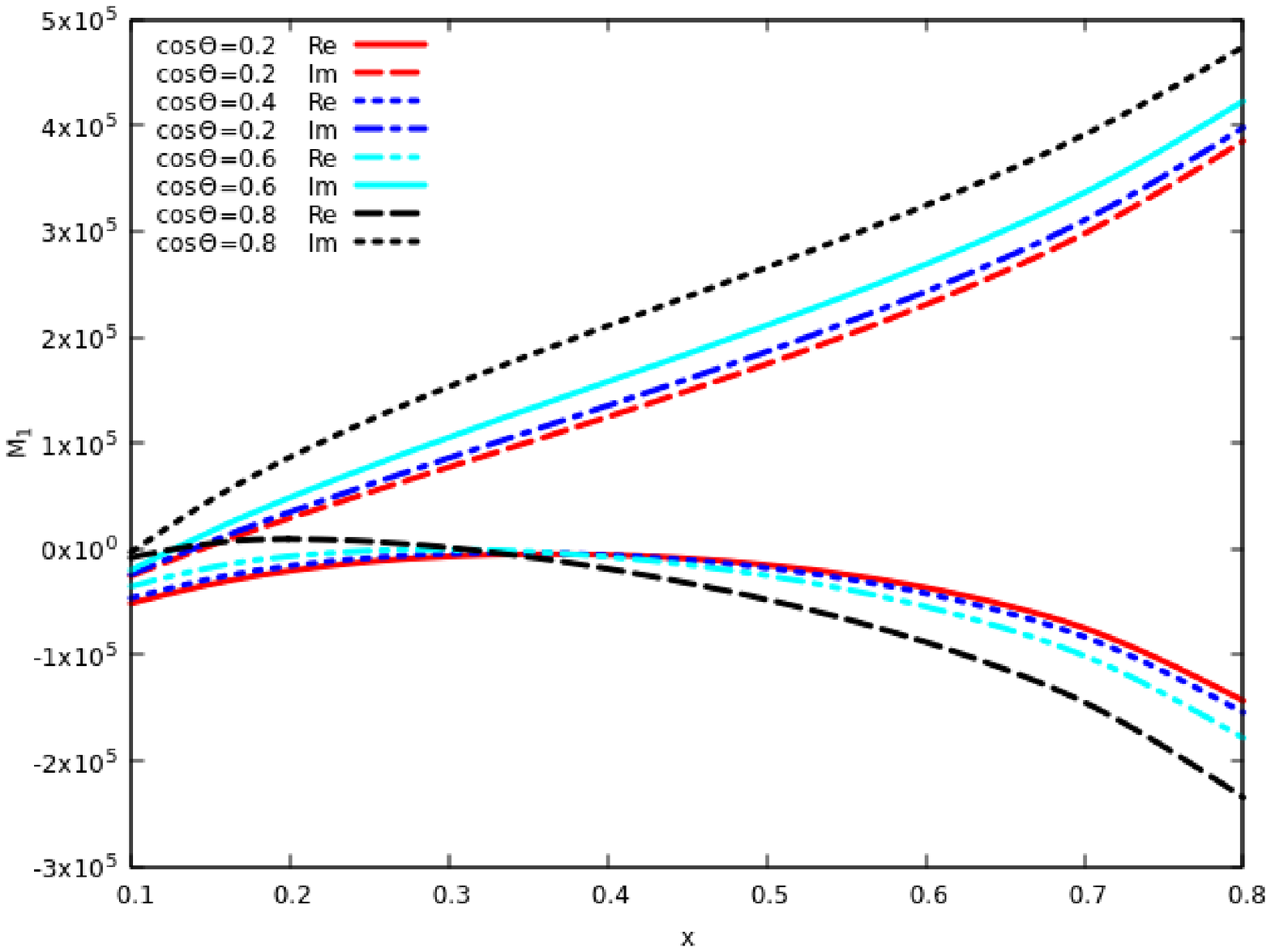}
\includegraphics[scale=0.45]{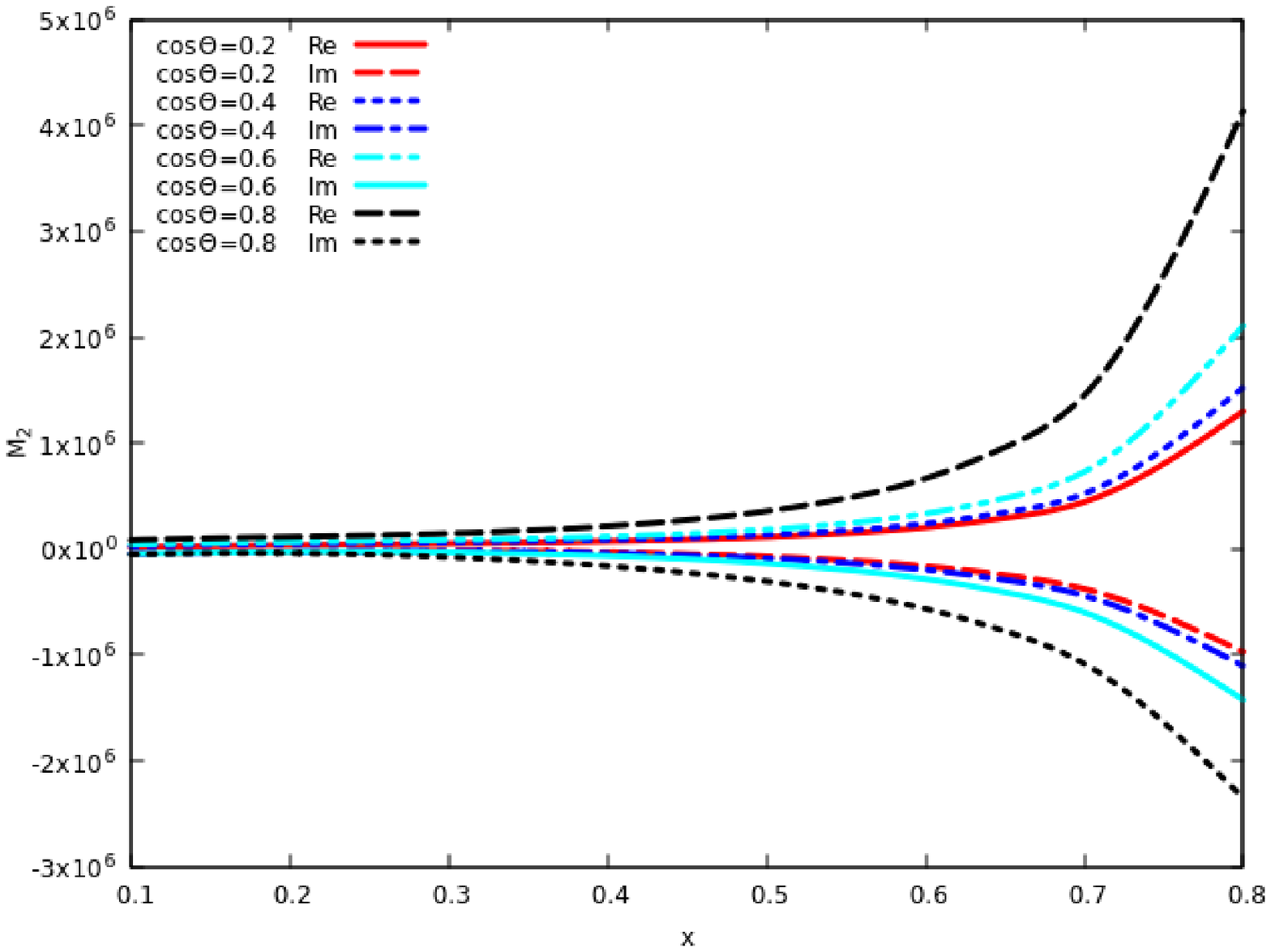}
\caption{Finite part of the two loop amplitudes for
projector 1,
$M_1= {\cal M}^{(2),fin}_{1,GG}/ {\cal M}^{(0)}_{1,GG}$ (left panel) and
projector 2,
$M_2= {\cal M}^{(2),fin}_{2,GG}/{\cal M}^{(0)}_{2,GG}$ (right panel),
suitably normalised,
have been plotted as a function of $x$, for different values of $\cos \theta$.
In the left panel, lines on the top half, corresponds to imaginary (Im) part
of the amplitude and in the lower half corresponds to the real (Re) part of the
amplitude.  The trend is reverse in the right panel for $M_2$
}
\label{fin}
\end{figure}

At one loop level, it is straight forward to show analytically that the IR poles are in
agreement with the predictions.  For the two loop case a fully analytical comparison was
possible only for poles $\epsilon^{-i}$ with $i=2-4$.  However, due to the large file
size for the $\epsilon^{-1}$ pole term, we made a comparison only at the numerical level.
We found full agreement with the predictions of Catani up to two loop level for all the
IR poles.  
Having obtained the IR pole cancellations, the finite part defined in eq.~\ref{mfin} can
be extracted by subtracting the IR poles.
Expressions for the finite part are too large to be presented here, 
however, they are being provided as ancillary files.
The finite part of the amplitude corresponding to the projectors 1 and 2 as a function
of $x$ for different values of $\cos \theta$ is plotted normalising appropriately,
in fig.~\ref{fin}.

The finite part ${\cal M}_{GJ}^{II(1),fin}$ in eq.~(\ref{OGGGJ}) starts at one loop and
contributes at order $a_s^4$ when combined with their respective Wilson coefficients.
For projector 1 and 2, we have
\begin{eqnarray}
{\cal M}_{1,GJ}^{II(1),fin} &=& - 24i ~m_A^{2} ~n_f ~\frac{(1+x)^2}{x} \,,
\\
{\cal M}_{2,GJ}^{II(1),fin} &=&-
12 i ~m_A^2~ n_f ~ \frac{(1-x^2)^2 (1+x^2) \sin^2 \theta}
{x \left((1+x^2)^2-(1-x^2)^2 \cos^2 \theta \right)} \,.
\end{eqnarray}

\section{Discussion and Conclusions}
In this paper, we have presented the two loop virtual amplitudes that are relevant for studying
production of pair of pseudo-scalar Higgs bosons in gluon fusion subprocess at the LHC.
This is the dominant sub process that is sensitive to its self coupling.  We have done this
computation in the EFT where top quark degrees of freedom is integrated out.
In the EFT, the pseudo-scalar Higgs boson directly couples to gluons and light quarks through
two local composite operators $O_G$ and $O_J$ respectively with the strengths proportional to
Wilson coefficients that are calculable in perturbative QCD.  We used dimensional regularisation
to regulate both UV and IR divergences.  The composite operators being CP odd, contain Levi-Civita
tensor and $\gamma_5$ which are inherently four dimensional objects.  Hence, a careful treatment was
needed to deal with them in $d$-dimensions.  We followed the prescription advocated by Larin.
This requires additional renormalisation for the singlet axial vector current up to two loops.
In addition, Larin's prescription requires finite renormalistion constant for
singlet axial current and is also available.  
Note that the composite operators mix under UV rerenormalistion.  
The corresponding renormalisation constants are already known and we use them to obtain UV finite 
two loop amplitudes.  
Unlike the amplitudes involving pair of Higgs bosons, we do not need any UV contact counter terms.
The UV finite amplitudes thus obtained contain IR divergences
due to the presence of massless partons in QCD.  We found that these IR poles are in agreement with the predictions
by Catani and it provides a test on the correctness of the computation.  Our results provide
one of the important components relevant for studies related to 
production of pair of pseudo-scalar Higgs bosons at the LHC up to order ${\cal O}(a_s^4)$.  
\label{sec:conc}
\\

\noindent
Acknowledgement:
We would like to acknowledge
the support of the CNRS LIA (Laboratoire International
Associe') THEP (Theoretical High Energy Physics) and
the INFRE-HEPNET (IndoFrench Network on High Energy Physics) of CEFIPRA/IFCPAR 
(Indo-French Centre for the Promotion of Advanced Research).  
We thank T. Gehrmann for providing us the master integrals and Roman Lee for his help with LiteRed.
We would like to thank Prasanna K Dhani for discussions.  AB would like to thank Pooja Mukherjee and
A.H. Ajjath for their guidance.  She would also like to thank her colleagues in the Theory 
division for their help.

\bibliographystyle{JHEP}
\bibliography{ggAA2L}

\providecommand{\href}[2]{#2}\begingroup\raggedright\begin{thebibliography}{10}

\bibitem{Aad:2012tfa}
{\scshape ATLAS} collaboration, G.~Aad et~al., \emph{{Observation of a new
  particle in the search for the Standard Model Higgs boson with the ATLAS
  detector at the LHC}},
  \href{https://doi.org/10.1016/j.physletb.2012.08.020}{\emph{Phys. Lett.}
  {\bfseries B716} (2012) 1} [\href{https://arxiv.org/abs/1207.7214}{{\ttfamily
  1207.7214}}].

\bibitem{Chatrchyan:2012xdj}
{\scshape CMS} collaboration, S.~Chatrchyan et~al., \emph{{Observation of a New
  Boson at a Mass of 125 GeV with the CMS Experiment at the LHC}},
  \href{https://doi.org/10.1016/j.physletb.2012.08.021}{\emph{Phys. Lett.}
  {\bfseries B716} (2012) 30}
  [\href{https://arxiv.org/abs/1207.7235}{{\ttfamily 1207.7235}}].

\bibitem{Fayet:1974pd}
P.~Fayet, \emph{{Supergauge Invariant Extension of the Higgs Mechanism and a
  Model for the electron and Its Neutrino}},
  \href{https://doi.org/10.1016/0550-3213(75)90636-7}{\emph{Nucl. Phys.}
  {\bfseries B90} (1975) 104}.

\bibitem{Fayet:1977yc}
P.~Fayet, \emph{{Spontaneously Broken Supersymmetric Theories of Weak,
  Electromagnetic and Strong Interactions}},
  \href{https://doi.org/10.1016/0370-2693(77)90852-8}{\emph{Phys. Lett.}
  {\bfseries 69B} (1977) 489}.

\bibitem{Dimopoulos:1981zb}
S.~Dimopoulos and H.~Georgi, \emph{{Softly Broken Supersymmetry and SU(5)}},
  \href{https://doi.org/10.1016/0550-3213(81)90522-8}{\emph{Nucl. Phys.}
  {\bfseries B193} (1981) 150}.

\bibitem{Sakai:1981gr}
N.~Sakai, \emph{{Naturalness in Supersymmetric Guts}},
  \href{https://doi.org/10.1007/BF01573998}{\emph{Z. Phys.} {\bfseries C11}
  (1981) 153}.

\bibitem{Inoue:1982pi}
K.~Inoue, A.~Kakuto, H.~Komatsu and S.~Takeshita, \emph{{Aspects of Grand
  Unified Models with Softly Broken Supersymmetry}},
  \href{https://doi.org/10.1143/PTP.68.927}{\emph{Prog. Theor. Phys.}
  {\bfseries 68} (1982) 927}.

\bibitem{Inoue:1983pp}
K.~Inoue, A.~Kakuto, H.~Komatsu and S.~Takeshita, \emph{{Renormalization of
  Supersymmetry Breaking Parameters Revisited}},
  \href{https://doi.org/10.1143/PTP.71.413}{\emph{Prog. Theor. Phys.}
  {\bfseries 71} (1984) 413}.

\bibitem{Inoue:1982ej}
K.~Inoue, A.~Kakuto, H.~Komatsu and S.~Takeshita, \emph{{Low-Energy Parameters
  and Particle Masses in a Supersymmetric Grand Unified Model}},
  \href{https://doi.org/10.1143/PTP.67.1889}{\emph{Prog. Theor. Phys.}
  {\bfseries 67} (1982) 1889}.

\bibitem{Martin:2007pg}
S.~P. Martin, \emph{{Three-loop corrections to the lightest Higgs scalar boson
  mass in supersymmetry}},
  \href{https://doi.org/10.1103/PhysRevD.75.055005}{\emph{Phys. Rev.}
  {\bfseries D75} (2007) 055005}
  [\href{https://arxiv.org/abs/hep-ph/0701051}{{\ttfamily hep-ph/0701051}}].

\bibitem{Harlander:2008ju}
R.~V. Harlander, P.~Kant, L.~Mihaila and M.~Steinhauser, \emph{{Higgs boson
  mass in supersymmetry to three loops}},
  \href{https://doi.org/10.1103/PhysRevLett.101.039901,
  10.1103/PhysRevLett.100.191602}{\emph{Phys. Rev. Lett.} {\bfseries 100}
  (2008) 191602} [\href{https://arxiv.org/abs/0803.0672}{{\ttfamily
  0803.0672}}].

\bibitem{Kant:2010tf}
P.~Kant, R.~V. Harlander, L.~Mihaila and M.~Steinhauser, \emph{{Light MSSM
  Higgs boson mass to three-loop accuracy}},
  \href{https://doi.org/10.1007/JHEP08(2010)104}{\emph{JHEP} {\bfseries 08}
  (2010) 104} [\href{https://arxiv.org/abs/1005.5709}{{\ttfamily 1005.5709}}].

\bibitem{Plehn:1996wb}
T.~Plehn, M.~Spira and P.~M. Zerwas, \emph{{Pair production of neutral Higgs
  particles in gluon-gluon collisions}},
  \href{https://doi.org/10.1016/0550-3213(96)00418-X,
  10.1016/S0550-3213(98)00406-4}{\emph{Nucl. Phys.} {\bfseries B479} (1996) 46}
  [\href{https://arxiv.org/abs/hep-ph/9603205}{{\ttfamily hep-ph/9603205}}].

\bibitem{Dawson:1998py}
S.~Dawson, S.~Dittmaier and M.~Spira, \emph{{Neutral Higgs boson pair
  production at hadron colliders: QCD corrections}},
  \href{https://doi.org/10.1103/PhysRevD.58.115012}{\emph{Phys. Rev.}
  {\bfseries D58} (1998) 115012}
  [\href{https://arxiv.org/abs/hep-ph/9805244}{{\ttfamily hep-ph/9805244}}].

\bibitem{Harlander:2002vv}
R.~V. Harlander and W.~B. Kilgore, \emph{{Production of a pseudoscalar Higgs
  boson at hadron colliders at next-to-next-to leading order}},
  \href{https://doi.org/10.1088/1126-6708/2002/10/017}{\emph{JHEP} {\bfseries
  10} (2002) 017} [\href{https://arxiv.org/abs/hep-ph/0208096}{{\ttfamily
  hep-ph/0208096}}].

\bibitem{Anastasiou:2002wq}
C.~Anastasiou and K.~Melnikov, \emph{{Pseudoscalar Higgs boson production at
  hadron colliders in NNLO QCD}},
  \href{https://doi.org/10.1103/PhysRevD.67.037501}{\emph{Phys. Rev.}
  {\bfseries D67} (2003) 037501}
  [\href{https://arxiv.org/abs/hep-ph/0208115}{{\ttfamily hep-ph/0208115}}].

\bibitem{Ravindran:2003um}
V.~Ravindran, J.~Smith and W.~L. van Neerven, \emph{{NNLO corrections to the
  total cross-section for Higgs boson production in hadron hadron collisions}},
  \href{https://doi.org/10.1016/S0550-3213(03)00457-7}{\emph{Nucl. Phys.}
  {\bfseries B665} (2003) 325}
  [\href{https://arxiv.org/abs/hep-ph/0302135}{{\ttfamily hep-ph/0302135}}].

\bibitem{Ahmed:2015qpa}
T.~Ahmed, T.~Gehrmann, P.~Mathews, N.~Rana and V.~Ravindran,
  \emph{{Pseudo-scalar Form Factors at Three Loops in QCD}},
  \href{https://doi.org/10.1007/JHEP11(2015)169}{\emph{JHEP} {\bfseries 11}
  (2015) 169} [\href{https://arxiv.org/abs/1510.01715}{{\ttfamily
  1510.01715}}].

\bibitem{Ravindran:2005vv}
V.~Ravindran, \emph{{On Sudakov and soft resummations in QCD}},
  \href{https://doi.org/10.1016/j.nuclphysb.2006.04.008}{\emph{Nucl. Phys.}
  {\bfseries B746} (2006) 58}
  [\href{https://arxiv.org/abs/hep-ph/0512249}{{\ttfamily hep-ph/0512249}}].

\bibitem{Ravindran:2006cg}
V.~Ravindran, \emph{{Higher-order threshold effects to inclusive processes in
  QCD}}, \href{https://doi.org/10.1016/j.nuclphysb.2006.06.025}{\emph{Nucl.
  Phys.} {\bfseries B752} (2006) 173}
  [\href{https://arxiv.org/abs/hep-ph/0603041}{{\ttfamily hep-ph/0603041}}].

\bibitem{Ahmed:2014cla}
T.~Ahmed, M.~Mahakhud, N.~Rana and V.~Ravindran, \emph{{Drell-Yan Production at
  Threshold to Third Order in QCD}},
  \href{https://doi.org/10.1103/PhysRevLett.113.112002}{\emph{Phys. Rev. Lett.}
  {\bfseries 113} (2014) 112002}
  [\href{https://arxiv.org/abs/1404.0366}{{\ttfamily 1404.0366}}].

\bibitem{Ahmed:2015qda}
T.~Ahmed, M.~C. Kumar, P.~Mathews, N.~Rana and V.~Ravindran,
  \emph{{Pseudo-scalar Higgs boson production at threshold N$^3$ LO and N$^3$
  LL QCD}}, \href{https://doi.org/10.1140/epjc/s10052-016-4199-1}{\emph{Eur.
  Phys. J.} {\bfseries C76} (2016) 355}
  [\href{https://arxiv.org/abs/1510.02235}{{\ttfamily 1510.02235}}].

\bibitem{Ahmed:2016otz}
T.~Ahmed, M.~Bonvini, M.~C. Kumar, P.~Mathews, N.~Rana, V.~Ravindran et~al.,
  \emph{{Pseudo-scalar Higgs boson production at N$^3$ LO$_{\text {A}}$ +N$^3$
  LL $'$}}, \href{https://doi.org/10.1140/epjc/s10052-016-4510-1}{\emph{Eur.
  Phys. J.} {\bfseries C76} (2016) 663}
  [\href{https://arxiv.org/abs/1606.00837}{{\ttfamily 1606.00837}}].

\bibitem{Glover:1987nx}
E.~W.~N. Glover and J.~J. van~der Bij, \emph{{HIGGS BOSON PAIR PRODUCTION VIA
  GLUON FUSION}},
  \href{https://doi.org/10.1016/0550-3213(88)90083-1}{\emph{Nucl. Phys.}
  {\bfseries B309} (1988) 282}.

\bibitem{Grigo:2013rya}
J.~Grigo, J.~Hoff, K.~Melnikov and M.~Steinhauser, \emph{{On the Higgs boson
  pair production at the LHC}},
  \href{https://doi.org/10.1016/j.nuclphysb.2013.06.024}{\emph{Nucl. Phys.}
  {\bfseries B875} (2013) 1} [\href{https://arxiv.org/abs/1305.7340}{{\ttfamily
  1305.7340}}].

\bibitem{Frederix:2014hta}
R.~Frederix, S.~Frixione, V.~Hirschi, F.~Maltoni, O.~Mattelaer, P.~Torrielli
  et~al., \emph{{Higgs pair production at the LHC with NLO and parton-shower
  effects}}, \href{https://doi.org/10.1016/j.physletb.2014.03.026}{\emph{Phys.
  Lett.} {\bfseries B732} (2014) 142}
  [\href{https://arxiv.org/abs/1401.7340}{{\ttfamily 1401.7340}}].

\bibitem{Maltoni:2014eza}
F.~Maltoni, E.~Vryonidou and M.~Zaro, \emph{{Top-quark mass effects in double
  and triple Higgs production in gluon-gluon fusion at NLO}},
  \href{https://doi.org/10.1007/JHEP11(2014)079}{\emph{JHEP} {\bfseries 11}
  (2014) 079} [\href{https://arxiv.org/abs/1408.6542}{{\ttfamily 1408.6542}}].

\bibitem{Degrassi:2016vss}
G.~Degrassi, P.~P. Giardino and R.~Gröber, \emph{{On the two-loop virtual QCD
  corrections to Higgs boson pair production in the Standard Model}},
  \href{https://doi.org/10.1140/epjc/s10052-016-4256-9}{\emph{Eur. Phys. J.}
  {\bfseries C76} (2016) 411}
  [\href{https://arxiv.org/abs/1603.00385}{{\ttfamily 1603.00385}}].

\bibitem{Borowka:2016ehy}
S.~Borowka, N.~Greiner, G.~Heinrich, S.~P. Jones, M.~Kerner, J.~Schlenk et~al.,
  \emph{{Higgs Boson Pair Production in Gluon Fusion at Next-to-Leading Order
  with Full Top-Quark Mass Dependence}},
  \href{https://doi.org/10.1103/PhysRevLett.117.079901,
  10.1103/PhysRevLett.117.012001}{\emph{Phys. Rev. Lett.} {\bfseries 117}
  (2016) 012001} [\href{https://arxiv.org/abs/1604.06447}{{\ttfamily
  1604.06447}}].

\bibitem{Borowka:2016ypz}
S.~Borowka, N.~Greiner, G.~Heinrich, S.~P. Jones, M.~Kerner, J.~Schlenk et~al.,
  \emph{{Full top quark mass dependence in Higgs boson pair production at
  NLO}}, \href{https://doi.org/10.1007/JHEP10(2016)107}{\emph{JHEP} {\bfseries
  10} (2016) 107} [\href{https://arxiv.org/abs/1608.04798}{{\ttfamily
  1608.04798}}].

\bibitem{Chen:2019lzz}
L.-B. Chen, H.~T. Li, H.-S. Shao and J.~Wang, \emph{{Higgs boson pair
  production via gluon fusion at N$^3$LO in QCD}},
  \href{https://arxiv.org/abs/1909.06808}{{\ttfamily 1909.06808}}.

\bibitem{deFlorian:2013uza}
D.~de~Florian and J.~Mazzitelli, \emph{{Two-loop virtual corrections to Higgs
  pair production}},
  \href{https://doi.org/10.1016/j.physletb.2013.06.046}{\emph{Phys. Lett.}
  {\bfseries B724} (2013) 306}
  [\href{https://arxiv.org/abs/1305.5206}{{\ttfamily 1305.5206}}].

\bibitem{Grigo:2015dia}
J.~Grigo, J.~Hoff and M.~Steinhauser, \emph{{Higgs boson pair production: top
  quark mass effects at NLO and NNLO}},
  \href{https://doi.org/10.1016/j.nuclphysb.2015.09.012}{\emph{Nucl. Phys.}
  {\bfseries B900} (2015) 412}
  [\href{https://arxiv.org/abs/1508.00909}{{\ttfamily 1508.00909}}].

\bibitem{deFlorian:2013jea}
D.~de~Florian and J.~Mazzitelli, \emph{{Higgs Boson Pair Production at
  Next-to-Next-to-Leading Order in QCD}},
  \href{https://doi.org/10.1103/PhysRevLett.111.201801}{\emph{Phys. Rev. Lett.}
  {\bfseries 111} (2013) 201801}
  [\href{https://arxiv.org/abs/1309.6594}{{\ttfamily 1309.6594}}].

\bibitem{Banerjee:2018lfq}
P.~Banerjee, S.~Borowka, P.~K. Dhani, T.~Gehrmann and V.~Ravindran,
  \emph{{Two-loop massless QCD corrections to the $g + g \to H + H$ four-point
  amplitude}}, \href{https://doi.org/10.1007/JHEP11(2018)130}{\emph{JHEP}
  {\bfseries 11} (2018) 130}
  [\href{https://arxiv.org/abs/1809.05388}{{\ttfamily 1809.05388}}].

\bibitem{H:2018hqz}
A.~H. Ajjath, P.~Banerjee, A.~Chakraborty, P.~K. Dhani, P.~Mukherjee, N.~Rana
  et~al., \emph{{Higgs pair production from bottom quark annihilation to NNLO
  in QCD}}, \href{https://doi.org/10.1007/JHEP05(2019)030}{\emph{JHEP}
  {\bfseries 05} (2019) 030}
  [\href{https://arxiv.org/abs/1811.01853}{{\ttfamily 1811.01853}}].

\bibitem{Li:2013flc}
Q.~Li, Q.-S. Yan and X.~Zhao, \emph{{Higgs Pair Production: Improved
  Description by Matrix Element Matching}},
  \href{https://doi.org/10.1103/PhysRevD.89.033015}{\emph{Phys. Rev.}
  {\bfseries D89} (2014) 033015}
  [\href{https://arxiv.org/abs/1312.3830}{{\ttfamily 1312.3830}}].

\bibitem{Maierhofer:2013sha}
P.~Maierhöfer and A.~Papaefstathiou, \emph{{Higgs Boson pair production merged
  to one jet}}, \href{https://doi.org/10.1007/JHEP03(2014)126}{\emph{JHEP}
  {\bfseries 03} (2014) 126} [\href{https://arxiv.org/abs/1401.0007}{{\ttfamily
  1401.0007}}].

\bibitem{deFlorian:2015moa}
D.~de~Florian and J.~Mazzitelli, \emph{{Higgs pair production at
  next-to-next-to-leading logarithmic accuracy at the LHC}},
  \href{https://doi.org/10.1007/JHEP09(2015)053}{\emph{JHEP} {\bfseries 09}
  (2015) 053} [\href{https://arxiv.org/abs/1505.07122}{{\ttfamily
  1505.07122}}].

\bibitem{Chetyrkin:1998mw}
K.~G. Chetyrkin, B.~A. Kniehl, M.~Steinhauser and W.~A. Bardeen,
  \emph{{Effective QCD interactions of CP odd Higgs bosons at three loops}},
  \href{https://doi.org/10.1016/S0550-3213(98)00594-X}{\emph{Nucl. Phys.}
  {\bfseries B535} (1998) 3}
  [\href{https://arxiv.org/abs/hep-ph/9807241}{{\ttfamily hep-ph/9807241}}].

\bibitem{Adler:1969gk}
S.~L. Adler, \emph{{Axial vector vertex in spinor electrodynamics}},
  \href{https://doi.org/10.1103/PhysRev.177.2426}{\emph{Phys. Rev.} {\bfseries
  177} (1969) 2426}.

\bibitem{Baikov:2009bg}
P.~A. Baikov, K.~G. Chetyrkin, A.~V. Smirnov, V.~A. Smirnov and M.~Steinhauser,
  \emph{{Quark and gluon form factors to three loops}},
  \href{https://doi.org/10.1103/PhysRevLett.102.212002}{\emph{Phys. Rev. Lett.}
  {\bfseries 102} (2009) 212002}
  [\href{https://arxiv.org/abs/0902.3519}{{\ttfamily 0902.3519}}].

\bibitem{Gehrmann:2010ue}
T.~Gehrmann, E.~W.~N. Glover, T.~Huber, N.~Ikizlerli and C.~Studerus,
  \emph{{Calculation of the quark and gluon form factors to three loops in
  QCD}}, \href{https://doi.org/10.1007/JHEP06(2010)094}{\emph{JHEP} {\bfseries
  06} (2010) 094} [\href{https://arxiv.org/abs/1004.3653}{{\ttfamily
  1004.3653}}].

\bibitem{tHooft:1972tcz}
G.~'t~Hooft and M.~J.~G. Veltman, \emph{{Regularization and Renormalization of
  Gauge Fields}},
  \href{https://doi.org/10.1016/0550-3213(72)90279-9}{\emph{Nucl. Phys.}
  {\bfseries B44} (1972) 189}.

\bibitem{Korner:1991sx}
J.~G. Korner, D.~Kreimer and K.~Schilcher, \emph{{A Practicable gamma(5) scheme
  in dimensional regularization}},
  \href{https://doi.org/10.1007/BF01559471}{\emph{Z. Phys.} {\bfseries C54}
  (1992) 503}.

\bibitem{Larin:1993tq}
S.~A. Larin, \emph{{The Renormalization of the axial anomaly in dimensional
  regularization}},
  \href{https://doi.org/10.1016/0370-2693(93)90053-K}{\emph{Phys. Lett.}
  {\bfseries B303} (1993) 113}
  [\href{https://arxiv.org/abs/hep-ph/9302240}{{\ttfamily hep-ph/9302240}}].

\bibitem{Larin:1991tj}
S.~A. Larin and J.~A.~M. Vermaseren, \emph{{The alpha-s**3 corrections to the
  Bjorken sum rule for polarized electroproduction and to the Gross-Llewellyn
  Smith sum rule}},
  \href{https://doi.org/10.1016/0370-2693(91)90839-I}{\emph{Phys. Lett.}
  {\bfseries B259} (1991) 345}.

\bibitem{Tarasov:1980au}
O.~V. Tarasov, A.~A. Vladimirov and A.~{\relax Yu}. Zharkov, \emph{{The
  Gell-Mann-Low Function of QCD in the Three Loop Approximation}},
  \href{https://doi.org/10.1016/0370-2693(80)90358-5}{\emph{Phys. Lett.}
  {\bfseries 93B} (1980) 429}.

\bibitem{Adler:1969er}
S.~L. Adler and W.~A. Bardeen, \emph{{Absence of higher order corrections in
  the anomalous axial vector divergence equation}},
  \href{https://doi.org/10.1103/PhysRev.182.1517}{\emph{Phys. Rev.} {\bfseries
  182} (1969) 1517}.

\bibitem{Zoller:2013ixa}
M.~F. Zoller, \emph{{OPE of the pseudoscalar gluonium correlator in massless
  QCD to three-loop order}},
  \href{https://doi.org/10.1007/JHEP07(2013)040}{\emph{JHEP} {\bfseries 07}
  (2013) 040} [\href{https://arxiv.org/abs/1304.2232}{{\ttfamily 1304.2232}}].

\bibitem{Kataev:1981aw}
A.~L. Kataev, N.~V. Krasnikov and A.~A. Pivovarov, \emph{{The Connection
  Between the Scales of the Gluon and Quark Worlds in Perturbative {QCD}}},
  \href{https://doi.org/10.1016/0370-2693(81)91161-8}{\emph{Phys. Lett.}
  {\bfseries 107B} (1981) 115}.

\bibitem{Kataev:1981gr}
A.~L. Kataev, N.~V. Krasnikov and A.~A. Pivovarov, \emph{{Two Loop Calculations
  for the Propagators of Gluonic Currents}},
  \href{https://doi.org/10.1016/0550-3213(82)90338-8,
  10.1016/S0550-3213(97)00101-6}{\emph{Nucl. Phys.} {\bfseries B198} (1982)
  508} [\href{https://arxiv.org/abs/hep-ph/9612326}{{\ttfamily
  hep-ph/9612326}}].

\bibitem{Nogueira:1991ex}
P.~Nogueira, \emph{{Automatic Feynman graph generation}},
  \href{https://doi.org/10.1006/jcph.1993.1074}{\emph{J. Comput. Phys.}
  {\bfseries 105} (1993) 279}.

\bibitem{Vermaseren:2000nd}
J.~A.~M. Vermaseren, \emph{{New features of FORM}},
  \href{https://arxiv.org/abs/math-ph/0010025}{{\ttfamily math-ph/0010025}}.

\bibitem{vonManteuffel:2012np}
A.~von Manteuffel and C.~Studerus, \emph{{Reduze 2 - Distributed Feynman
  Integral Reduction}},  \href{https://arxiv.org/abs/1201.4330}{{\ttfamily
  1201.4330}}.

\bibitem{Gehrmann:2013cxs}
T.~Gehrmann, L.~Tancredi and E.~Weihs, \emph{{Two-loop master integrals for $q
  \bar{q} \to VV$: the planar topologies}},
  \href{https://doi.org/10.1007/JHEP08(2013)070}{\emph{JHEP} {\bfseries 08}
  (2013) 070} [\href{https://arxiv.org/abs/1306.6344}{{\ttfamily 1306.6344}}].

\bibitem{Gehrmann:2014bfa}
T.~Gehrmann, A.~von Manteuffel, L.~Tancredi and E.~Weihs, \emph{{The two-loop
  master integrals for $q\overline{q} \to VV$}},
  \href{https://doi.org/10.1007/JHEP06(2014)032}{\emph{JHEP} {\bfseries 06}
  (2014) 032} [\href{https://arxiv.org/abs/1404.4853}{{\ttfamily 1404.4853}}].

\bibitem{Tkachov:1981wb}
F.~V. Tkachov, \emph{{A Theorem on Analytical Calculability of Four Loop
  Renormalization Group Functions}},
  \href{https://doi.org/10.1016/0370-2693(81)90288-4}{\emph{Phys. Lett.}
  {\bfseries 100B} (1981) 65}.

\bibitem{Chetyrkin:1981qh}
K.~G. Chetyrkin and F.~V. Tkachov, \emph{{Integration by Parts: The Algorithm
  to Calculate beta Functions in 4 Loops}},
  \href{https://doi.org/10.1016/0550-3213(81)90199-1}{\emph{Nucl. Phys.}
  {\bfseries B192} (1981) 159}.

\bibitem{Gehrmann:1999as}
T.~Gehrmann and E.~Remiddi, \emph{{Differential equations for two loop four
  point functions}},
  \href{https://doi.org/10.1016/S0550-3213(00)00223-6}{\emph{Nucl. Phys.}
  {\bfseries B580} (2000) 485}
  [\href{https://arxiv.org/abs/hep-ph/9912329}{{\ttfamily hep-ph/9912329}}].

\bibitem{Anastasiou:2004vj}
C.~Anastasiou and A.~Lazopoulos, \emph{{Automatic integral reduction for higher
  order perturbative calculations}},
  \href{https://doi.org/10.1088/1126-6708/2004/07/046}{\emph{JHEP} {\bfseries
  07} (2004) 046} [\href{https://arxiv.org/abs/hep-ph/0404258}{{\ttfamily
  hep-ph/0404258}}].

\bibitem{Smirnov:2008iw}
A.~V. Smirnov, \emph{{Algorithm FIRE -- Feynman Integral REduction}},
  \href{https://doi.org/10.1088/1126-6708/2008/10/107}{\emph{JHEP} {\bfseries
  10} (2008) 107} [\href{https://arxiv.org/abs/0807.3243}{{\ttfamily
  0807.3243}}].

\bibitem{Studerus:2009ye}
C.~Studerus, \emph{{Reduze-Feynman Integral Reduction in C++}},
  \href{https://doi.org/10.1016/j.cpc.2010.03.012}{\emph{Comput. Phys. Commun.}
  {\bfseries 181} (2010) 1293}
  [\href{https://arxiv.org/abs/0912.2546}{{\ttfamily 0912.2546}}].

\bibitem{Lee:2013mka}
R.~N. Lee, \emph{{LiteRed 1.4: a powerful tool for reduction of multiloop
  integrals}}, \href{https://doi.org/10.1088/1742-6596/523/1/012059}{\emph{J.
  Phys. Conf. Ser.} {\bfseries 523} (2014) 012059}
  [\href{https://arxiv.org/abs/1310.1145}{{\ttfamily 1310.1145}}].

\bibitem{Laporta:2001dd}
S.~Laporta, \emph{{High precision calculation of multiloop Feynman integrals by
  difference equations}}, \href{https://doi.org/10.1016/S0217-751X(00)00215-7,
  10.1142/S0217751X00002157}{\emph{Int. J. Mod. Phys.} {\bfseries A15} (2000)
  5087} [\href{https://arxiv.org/abs/hep-ph/0102033}{{\ttfamily
  hep-ph/0102033}}].

\bibitem{Catani:1998bh}
S.~Catani, \emph{{The Singular behavior of QCD amplitudes at two loop order}},
  \href{https://doi.org/10.1016/S0370-2693(98)00332-3}{\emph{Phys. Lett.}
  {\bfseries B427} (1998) 161}
  [\href{https://arxiv.org/abs/hep-ph/9802439}{{\ttfamily hep-ph/9802439}}].

\bibitem{Sterman:2002qn}
G.~F. Sterman and M.~E. Tejeda-Yeomans, \emph{{Multiloop amplitudes and
  resummation}},
  \href{https://doi.org/10.1016/S0370-2693(02)03100-3}{\emph{Phys. Lett.}
  {\bfseries B552} (2003) 48}
  [\href{https://arxiv.org/abs/hep-ph/0210130}{{\ttfamily hep-ph/0210130}}].

\bibitem{Becher:2009cu}
T.~Becher and M.~Neubert, \emph{{Infrared singularities of scattering
  amplitudes in perturbative QCD}},
  \href{https://doi.org/10.1103/PhysRevLett.102.162001,
  10.1103/PhysRevLett.111.199905}{\emph{Phys. Rev. Lett.} {\bfseries 102}
  (2009) 162001} [\href{https://arxiv.org/abs/0901.0722}{{\ttfamily
  0901.0722}}].

\bibitem{Gardi:2009qi}
E.~Gardi and L.~Magnea, \emph{{Factorization constraints for soft anomalous
  dimensions in QCD scattering amplitudes}},
  \href{https://doi.org/10.1088/1126-6708/2009/03/079}{\emph{JHEP} {\bfseries
  03} (2009) 079} [\href{https://arxiv.org/abs/0901.1091}{{\ttfamily
  0901.1091}}].

\end{thebibliography}\endgroup
\end{document}